\documentclass[twocolumn,preprintnumbers,amsmath,amssymb]{revtex4}
\usepackage{siunitx}
\usepackage{graphicx}
\usepackage{dcolumn}
\usepackage{bm}
\begin{document}

\preprint{}

\title{Charging up the cold: Formation of doubly- and triply-charged fullerene dimers in superfluid helium nanodroplets}

\author{Lisa Ganner}%
\author{Stefan Bergmeister}%
\author{Lucas Lorenz}%
\author{Milan Ončák}%
\author{Paul Scheier}%
\author{Elisabeth Gruber}
\affiliation{Institute for Ion Physics and Applied Physics, University of Innsbruck, Technikerstra{\ss}e 25, 6020 Innsbruck, Austria}

\date{\today}

\begin{abstract}
Sequential ionization of fullerene cluster ions (C$_{60}$)$_{n}^{+}$ within multiply-charged helium nanodroplets leads to the intriguing phenomenon of forming and stabilizing doubly- and triply-charged fullerene oligomers. Surprisingly, we have detected (C$_{60}$)$_{2}^{2+}$ and (C$_{60}$)$_{2}^{3+}$, indicating that dimers, rather than the previously established pentamers and dodecamers, are the smallest fullerene cluster sizes  capable of stabilizing two and even three charges. This remarkable resilience against Coulomb explosion is achieved through efficient cooling within the superfluid environment of helium nanodroplets, and a sequential ionization scheme that populates covalently bound or physisorbed fullerene dimers. Calculations support the stability of four differently bonded (C$_{60}$)$_{2}^{2+}$ and (C$_{60}$)$_{2}^{3+}$ isomers and predict a low Coulomb barrier (\textless0.4 eV) preventing even dissociation of cold van der Waals complexes.

\end{abstract}

\maketitle


Fullerenes and fullerene clusters, owing to their distinctive electronic and geometric configurations, exhibit a wide range of structural, electronic, magnetic, and chemical properties, making them highly promising for nano-technological applications. When a sufficient amount of energy is imparted to the loosely bound van der Waals (vdW) C$_{60}$ clusters, it leads to the opening of the cage structure and the formation of new chemical bonds, resulting in the creation of dimers, oligomers, or even polymers. These bonds can manifest as either sp$^{3}$ bonds (single bonds or 2+2 cycloaddition bonds) or sp$^{2}$ bonds.

In laboratory settings, the polymerization of C$_{60}$ has been achieved through the application of high pressure and temperature~\cite{erohin2022} or by irradiating C$_{60}$ films with visible or ultraviolet light~\cite{rao1993}. Dimerization and polymerization have also been observed when subjecting C$_{60}$ to swift heavy ion irradiation on thin Si films and quartz substrates~\cite{bajwa2003}. Furthermore, fullerene-coalescence of C$_{60}$ within single-wall nanotubes has been observed through heat treatment~\cite{bandow2001} or electron bombardment~\cite{koshino2010}.

In the gas phase, the fusion of individual fullerene molecules within weakly-bound clusters to form larger fullerene polymers, denoted as (C$_{60}$)$_{n}^{+}$, has been observed under specific conditions, such as ultra-short pulse laser irradiation~\cite{heden2005, campbell2006} and deep inelastic scattering collisions involving C$_{60}^+$ + C$_{60}$~\cite{campbell1993,glotov2000, rohmund1996_1,rohmund1996_2}. Additionally, studies employing keV alpha particles colliding with C$_{60}$ vdW clusters have revealed the formation of C$_{59}^+$ and C$_{58}^+$, which subsequently react rapidly with C$_{60}$ to yield C$_{119}^+$ and C$_{118}^+$, respectively~\cite{zettergren2013}.

In addition to investigating the dimerization and polymerization of fullerenes, the ionization tolerance of the C$_{60}$ monomer as well as of C$_{60}$ oligomers has been studied. According to calculations, highly-charged ions, up to C$_{60}^{26+}$, can be formed without the destruction of the fullerene cage due to Coulomb explosion~\cite{sadjadi2021}. When dealing with fullerene clusters denoted as (C$_{60}$)$_{n}^{q+}$, critical cluster sizes of \textit{n} = 5, 10, 21 and 33 have been experimentally identified for different charge states $q$, specifically 2, 3, 4 and 5, respectively~\cite{manil2003}. These measurements were conducted by producing multiply-charged fullerene clusters in collisions with highly-charged Xe$^{20+}$ and Xe$^{30+}$ ions, and their stabilities and fragmentation patterns were analyzed using high-resolution time-of-flight mass-spectrometry. Furthermore, the stability of multiply-charged fullerene vdW cluster ions has also been examined theoretically, utilizing a simple model based on classical electrostatics. This analysis has shown that the pentamer represents the smallest stable dication~\cite{huber2016}. However, it is worth  noting that the authors have indicated that kinetic barriers theoretically allow for the stabilization of even smaller fullerene vdW cluster ions, such as dimers, although achieving this would require highly efficient methods of active cooling.

\begin{figure}
\centering
\includegraphics[width=1\linewidth]{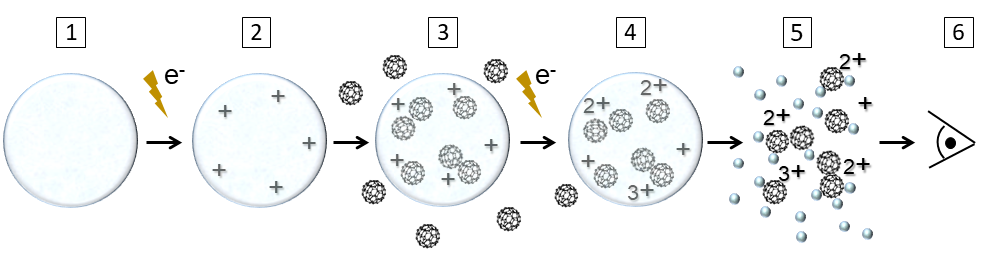}
\caption{Schematic illustration of the formation of doubly- and triply-charged C$_{60}$ cluster ions through sequential ionization within the sub-Kelvin environment of HNDs. (1) Formation of HNDs via supersonic jet expansion. (2) Electron impact ionization of the HNDs. (3) Introduction of C$_{60}$ molecules into the gas phase through sublimation in vacuum, subsequently doping the multiply-charged HNDs. (4) Second electron impact ionization of the charged and doped HNDs. This process involves sequential Penning ionization, leading to the creation of doubly- and triply-charged C$_{60}$ cluster ions. (5) Gentle collisions with helium gas at room temperature liberate the dopant cluster ions from the HND environment. (6) Detection is performed using a high-resolution time-of-flight mass-spectrometer.}
\label{fig1} 
\end{figure}

In this contribution, we present a novel experimental technique which allows the formation and stabilization of multiply-charged dopant ions within the sub-Kelvin environment of helium nanodroplets (HNDs). We apply the method on fullerene clusters and present the unexpected result of the formation and stabilization of doubly- and even triply-charged C$_{60}$ dimers. We discuss sequential Penning ionization as the likely mechanism by which doubly- and triply-charged ions are formed in doped HNDs. Additionally, we investigate the properties of (C$_{60}$)$_2^{2+}$ and (C$_{60}$)$_2^{3+}$ employing density functional theory calculations (BP86-D3/def2SVP as implemented in Gaussian; see the Supplemental Material for details)~\cite{grimme2010,g16}, showing that a fission barrier is sufficient to stabilize the doubly- and triply-charged fullerene dimers once trapped in their respective ground state configurations via efficient cooling by the surrounding He matrix. 

In our experimental setup (see FIG.~\ref{fig1}), we generate HNDs through supersonic jet expansion of high-pressure (22~bar) and pre-cooled helium gas through a \SI{5}{\mathrm{\mu}\meter} nozzle cooled to \SI{8.5}{K} in an ultra-high vacuum environment. Under these conditions, the produced HNDs contain at average \SI{4e6} He atoms ~\cite{toennies2004}. After the formation, the HNDs become highly-charged~\cite{laimer2019} by multiple electron impact ionization. The electron gun for electron ionization (EI) is operated at \SI{40}{V} acceleration voltage and electron currents of \SI{300}{\mathrm{\mu}A}.
At electron energies above the helium ionization threshold of 24.6~eV, besides electronic excitation forming He$^{*}$, the incident electron can also directly ionize helium atoms forming He$^{+}$, which will terminate in the formation of He$_{2}^{+}$~\cite{mauracher2018}. He$^{+}$, or He$_{2}^{+}$, respectively, might also be formed via Penning ionization of two metastable He atoms, He$^{*}$ + He$^{*}$ $\rightarrow$ He$^{+}$ + He + $\Delta$$E$. Due to Coloumb repulsion, the formed charge centers He$^{+}$ or He$_{2}^{+}$ migrate to the surface of the HNDs and arrange in a lattice-like structure~\cite{feinberg2022}. After the formation of multiply-charged HNDs, we select HNDs with a specific mass-per-charge ratio by using an electrostatic sector field. For the presented measurements, the settings of the sector field are chosen so that we select HNDs containing on average \SI{1.5e5} He atoms per charge with an average charge state of $\sim$33 (for the calculation of the average charge state see Supplemental Material).

In the next step, these multiply-charged HNDs are doped with C$_{60}$ molecules, which are brought into the gas phase by evaporation of C$_{60}$ powder (SES Research, 99.99$\%$) in a resistively heated oven in the pick-up chamber. These dopant molecules are attracted by the charges across the HND surface and are ionized through charge transfer processes, He$^{+}$ + C$_{60}$ $\rightarrow$ He + C$_{60}^{+}$ + $\Delta$$E$. The high potential energy of He$^{+}$ (24.6~eV) and of He$_{2}^{+}$ (22~eV) is sufficient to singly or even doubly ionize C$_{60}$. The ionization energies of the fullerene C$_{60}$ were determined in earlier measurements and found to be IE$_{1}=7.6$~eV, IE$_{2}=11.4$~eV and IE$_{3}=16.6$~eV~\cite{scheier1994}. Further pickup of C$_{60}$ molecules results in the formation of larger C$_{60}$ cluster ions, i.e., dimers, trimers or larger oligomers, which are quickly cooled to the equilibrium temperature of 0.37~K within the superfluid environment of HNDs. Any excess energy is dissipated into the surrounding helium matrix and released through evaporative cooling. The average of the formed cluster size is tuned by varying the C$_{60}$ pressure in the pick-up chamber with the oven temperature. 

Further exposure to an electron beam, using a second EI source (also operated at \SI{40}{V}, delivering \SI{300}{\mathrm{\mu}A} electron current), leads, once again, to the formation of He$^{+}$ as well as He$^{*}$. The formation of doubly- or triply-charged dopant ions by direct electron impact ionization is quite unlikely as the combined cross-section of the He atoms surrounding the charge centers is much larger than that of the dopant ions. While He$^{+}$ will not further interact with the charged dopant ions due to Coulomb repulsion, He$^{*}$ Penning ionizes (C$_{60}$)$_{n}^{+}$ to (C$_{60}$)$_{n}^{2+}$ or (C$_{60}$)$_{n}^{3+}$. Eventually, the ions are gently liberated from the HND by collisions with helium gas at room temperature and detected in a time-of-flight mass-spectrometer~\cite{bergmeister2023}. 

\begin{figure}
\centering
\includegraphics[width=1\linewidth]{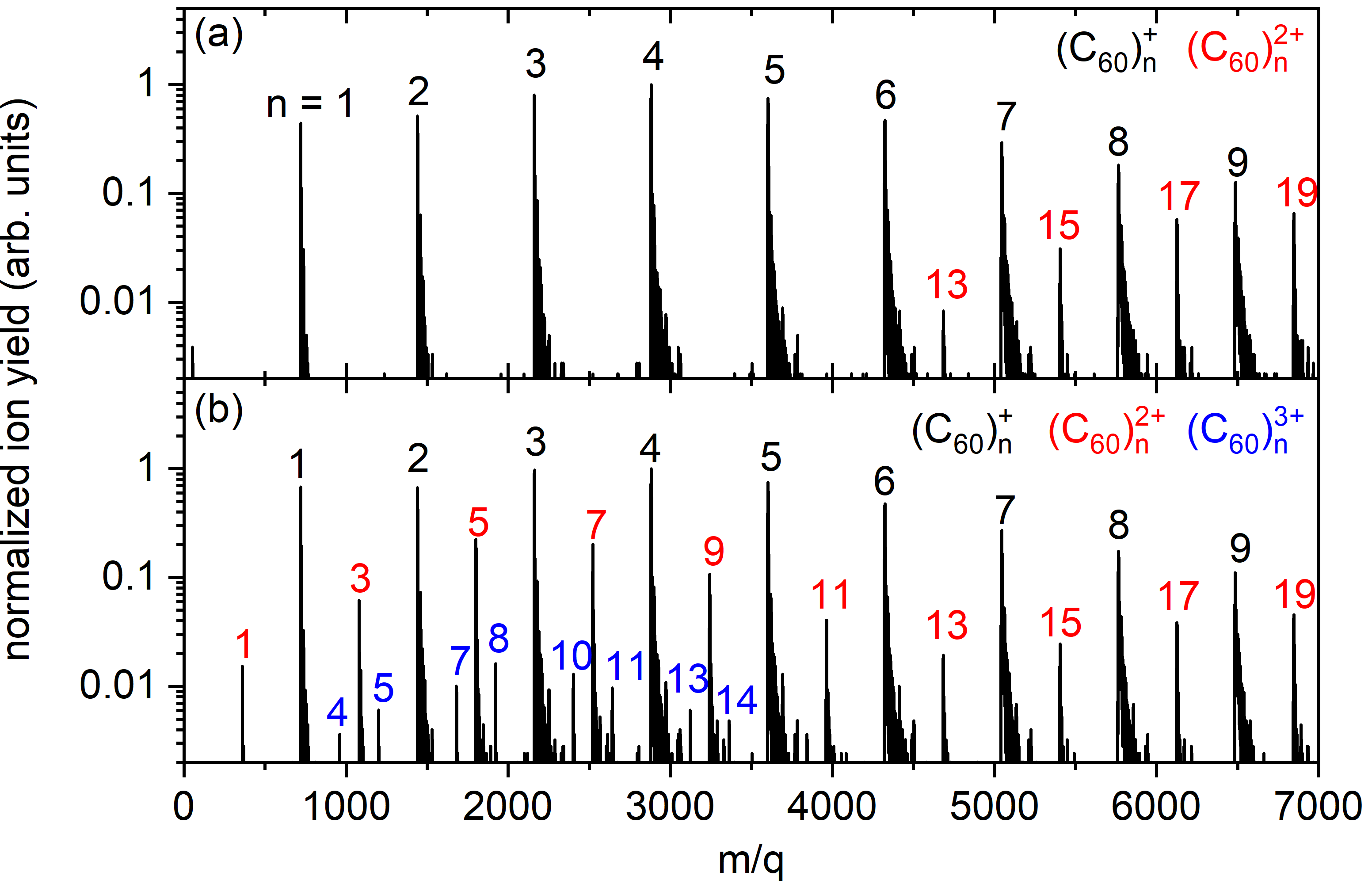}
\caption{\label{fig2} (a) Mass spectrum of C$_{60}$ cluster ions obtained when only the first EI source is activated. (b) Mass spectrum of C$_{60}$ cluster ions obtained when both the first and second EI sources are activated. Sequential Penning ionization results in the creation of doubly- and triply-charged C$_{60}$ cluster ions.}
\end{figure}

\begin{figure}
\centering
\includegraphics[width=1\linewidth]{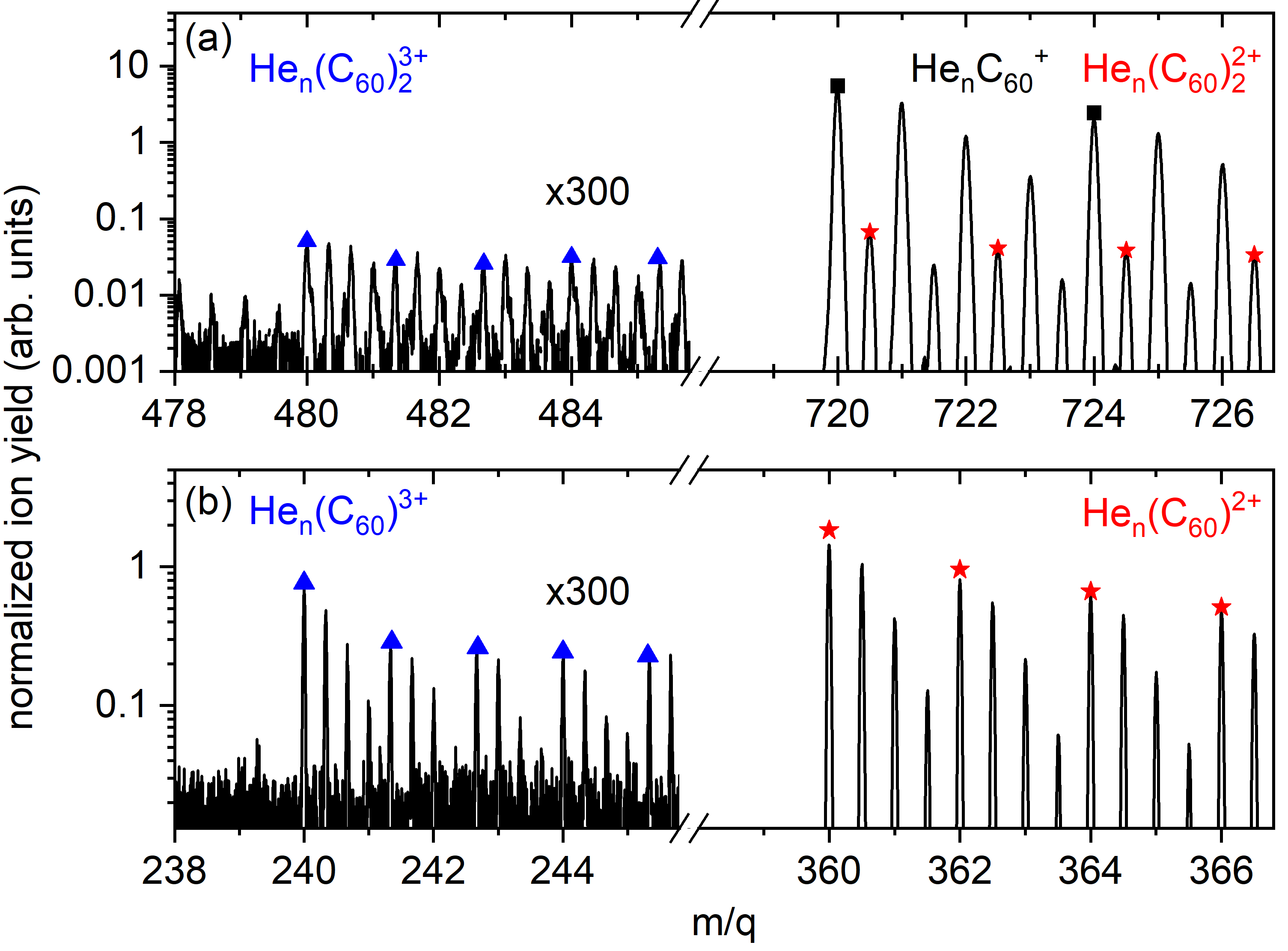}
\caption{\label{fig3} A closer examination of the lower mass range clearly reveals the presence of doubly- (labeled He$_{n}$(C$_{60}$)$_{2}^{2+}$ contain one $^{13}$C) and triply-charged C$_{60}$ dimers (a), as well as doubly- and triply-charged C$_{60}$ monomers (b). To enhance the visibility, the spectra corresponding to ions with higher charge states have been magnified by the labeled factor.}
\end{figure}

Figure~\ref{fig2}a displays the mass spectrum when only the first EI source is active. The spectrum is dominated by singly-charged C$_{60}$ cluster ions with a few He atoms attached. Beyond a mass of 4500~amu, doubly-charged C$_{60}$ cluster ions start to appear in addition to singly-charged ones. However, turning on the second EI source while maintaining all other parameters (voltages, pressures) results in a significantly different mass spectrum (FIG.~\ref{fig2}b); doubly- and triply-charged C$_{60}$ cluster ions become clearly visible. Surprisingly, even doubly- and triply-charged dimers are observed, as shown in more detail in FIG.~\ref{fig3}a. Although the mass-to-charge ratio of doubly-charged dimers coincides with that of singly-charged monomers, the presence of C$_{60}$ isotopologues allows for their unambiguous identification.

\begin{figure}
\centering
\includegraphics[width=1\linewidth]{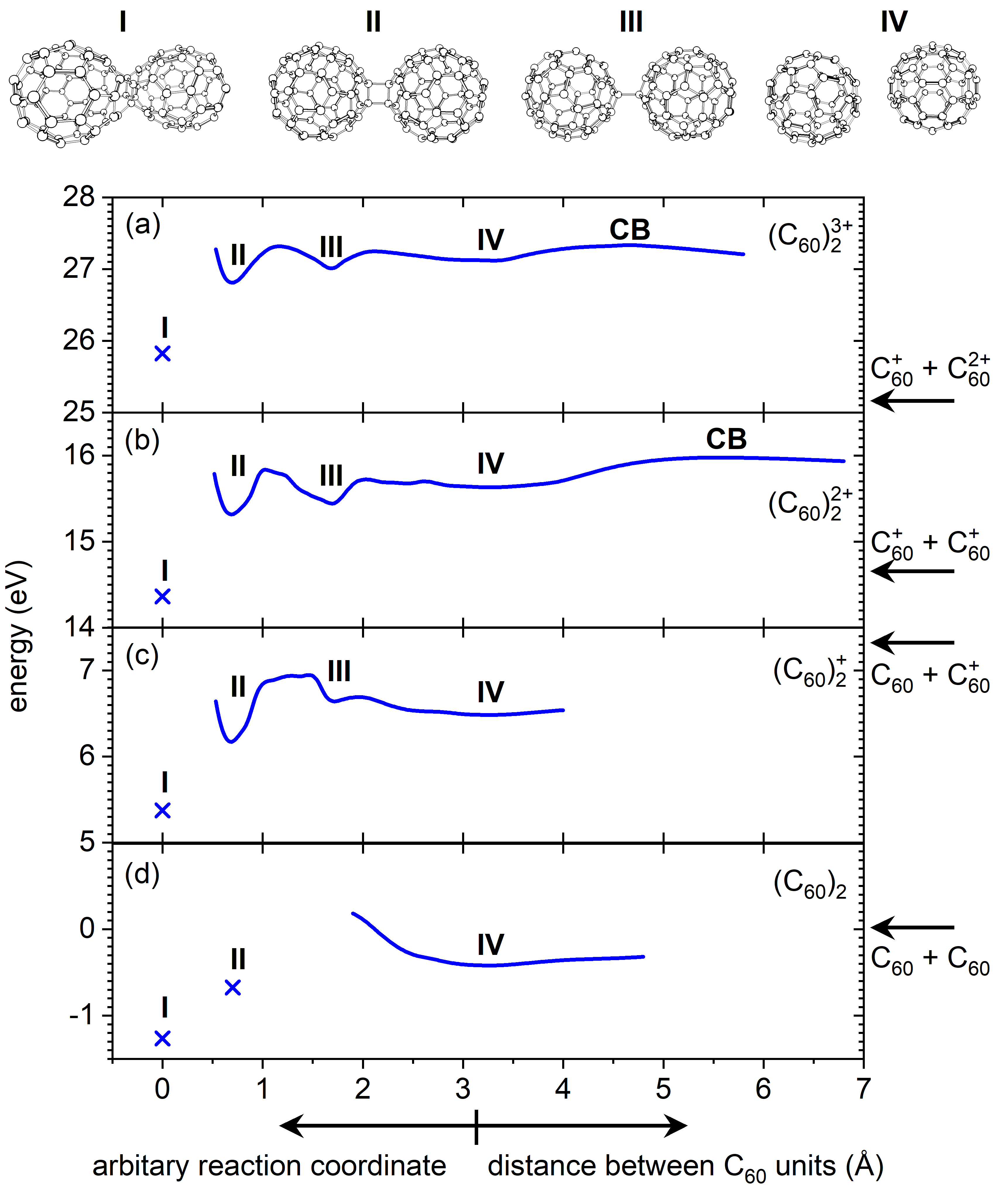}
\caption{\label{fig4} Potential energy scans connecting various isomers for neutral and charged (C$_{60}$)$_2$ complexes with respect to the 2 C$_{60}$ asymptote as calculated at the BP86-D3/def2SVP level of theory. Three covalently bound minima (I, II, III) and a physisorbed structure (IV) are considered, Coulomb barriers are denoted with CB. See the Supplemental Material for calculation details.}
\end{figure}

Our measurements unequivocally demonstrate that di- and even tricationic dimers can be stabilized, contradicting previous investigations~\cite{manil2003, huber2016}. To explain these differences, we examine the potential energy surfaces calculated for neutral, singly-, doubly- and triply-charged fullerene dimers of various configurations. Specifically, we consider structures with four (I), two (II) and one (III) covalent bond along with a physisorbed structure (IV) (see FIG.~\ref{fig4}). 
In our study, the process of forming multiply-charged fullerene dimers initiates with the ionization of a fullerene monomer through charge transfer with He$^{+}$ or He$_{2}^{+}$, close to the HND surface. The approach of a second neutral C$_{60}$ molecule via sequential pick-up leads to the formation of a singly-charged fullerene dimer (C$_{60}$)$_{2}^{+}$. Examining the potential energy curve in FIG.~\ref{fig4}c, starting on the C$_{60}$ + C$_{60}^{+}$ asymptote, reveals several reachable isomeric configurations, with minima separated by small barriers. 

The efficient cooling of (C$_{60}$)$_{2}^{+}$ within the helium matrix allows trapping in the local minima of the potential energy curve. Despite our current lack of knowledge regarding the exact structure of the formed cluster ions in HNDs, recent spectroscopic investigations of (C$_{60}$)$_2^{+}$  in the infrared wavelength range (1100$-$1600~cm$^{-1}$)~\cite{kappe2023} indicate the prevalence of peanut-shaped isomers with 4 covalent bonds (I) over other structures. However, it is crucial to note, that the formation of several isomers cannot be ruled out. Only the formation of cationic buckytubes can be excluded as their production requires the breaking and formation of numerous covalent bonds, and the spectroscopic investigations as well do not provide any evidence for their formation. Additionally, we performed collision-induced dissociation measurements of singly- and doubly-charged fullerene cluster ions (see Supplemental Material), revealing a clear preference for the loss of intact fullerene units, rather than other fragments. This result contrasts with what is expected in the case of buckytubes. It would be desirable to be able to measure which isomers are formed and with what efficiency, also in order to evaluate the cooling rate relative to the dimer formation rate. However, these measurements are not possible with the existing experimental setup and are beyond the scope of this work. 

Starting in the minima of the potential energy curve in FIG.~\ref{fig4}c, vertical ionization followed by efficient cooling will lead to the population of the minima of the potential energy curves for (C$_{60}$)$_{2}^{2+}$ (FIG.~\ref{fig4}b) and (C$_{60}$)$_{2}^{3+}$ (FIG.~\ref{fig4}a). Any excess energy, including kinetic and internal energy such as vibrational energy, is dissipated into the surrounding helium matrix. Experimentally, the post-ionization is achieved through Penning ionization with He$^{*}$, generated by the electron shower from the second EI source. The calculated upper bounds of the Coulomb barriers of about 0.4~eV and 0.2~eV in (C$_{60}$)$_{2}^{2+}$ and (C$_{60}$)$_{2}^{3+}$, respectively, prevent dissociation and explain the stability of the doubly- and triply-charged fullerene dimers. 

In previous studies where neutral C$_{60}$ vdW bound oligomers were highly-charged upon vertical ionization~\cite{manil2003, huber2016}, the starting point is the potential energy curve in FIG.~\ref{fig4}d. Upon ionization, possible changes in the fullerene cage leads to intramolecular vibration which when coupled to intermolecular vibration of the two fullerene cages shifts the population of vibrational states above the Coulomb barrier and, ultimately, to dissociation. Only the Coulomb barriers of the pentamer and dodecamer were found sufficiently high to stabilize two or three charges, respectively~\cite{manil2003, huber2016}.

We have shown, that our novel instrumental setup enables a precise sequence of ionization and cluster formation, accompanied by effective cooling in between. This process enhances the conformational flexibility of the formed ions, facilitating the stabilization of multiply-charged ions of very low stability. This advancement opens avenues for gaining deeper insight into the formation, stability, and spectroscopy of multiply-charged fullerene oligomers, along with other highly-charged, metastable systems.

\begin{acknowledgments}
This work was supported by the Austrian Science Fund FWF, Projects W1259, T1181, P31149, P34563. The computational results presented have been achieved using the HPC infrastructure LEO of the University of Innsbruck. This article is based upon work from COST Action CA21126 - Carbon molecular nanostructures in space (NanoSpace), supported by COST (European Cooperation in Science and Technology).
\end{acknowledgments}

\bibliographystyle{apsrev}
\bibliography{literature}

\end{document}


\title{Supplemental Material: Charging up the cold: Formation of doubly- and triply-charged fullerene dimers in superfluid helium nanodroplets}

\author{Lisa Ganner, Stefan Bergmeister, Lucas Lorenz, Milan On\v{c}\'{a}k, Paul Scheier, Elisabeth Gruber}%

\maketitle

\section{Calculation of the average charge state of the mass-per-charge selected HND ions} 

\begin{figure}
\begin{minipage}{0.48\textwidth}
\centering
\includegraphics[width=8cm]{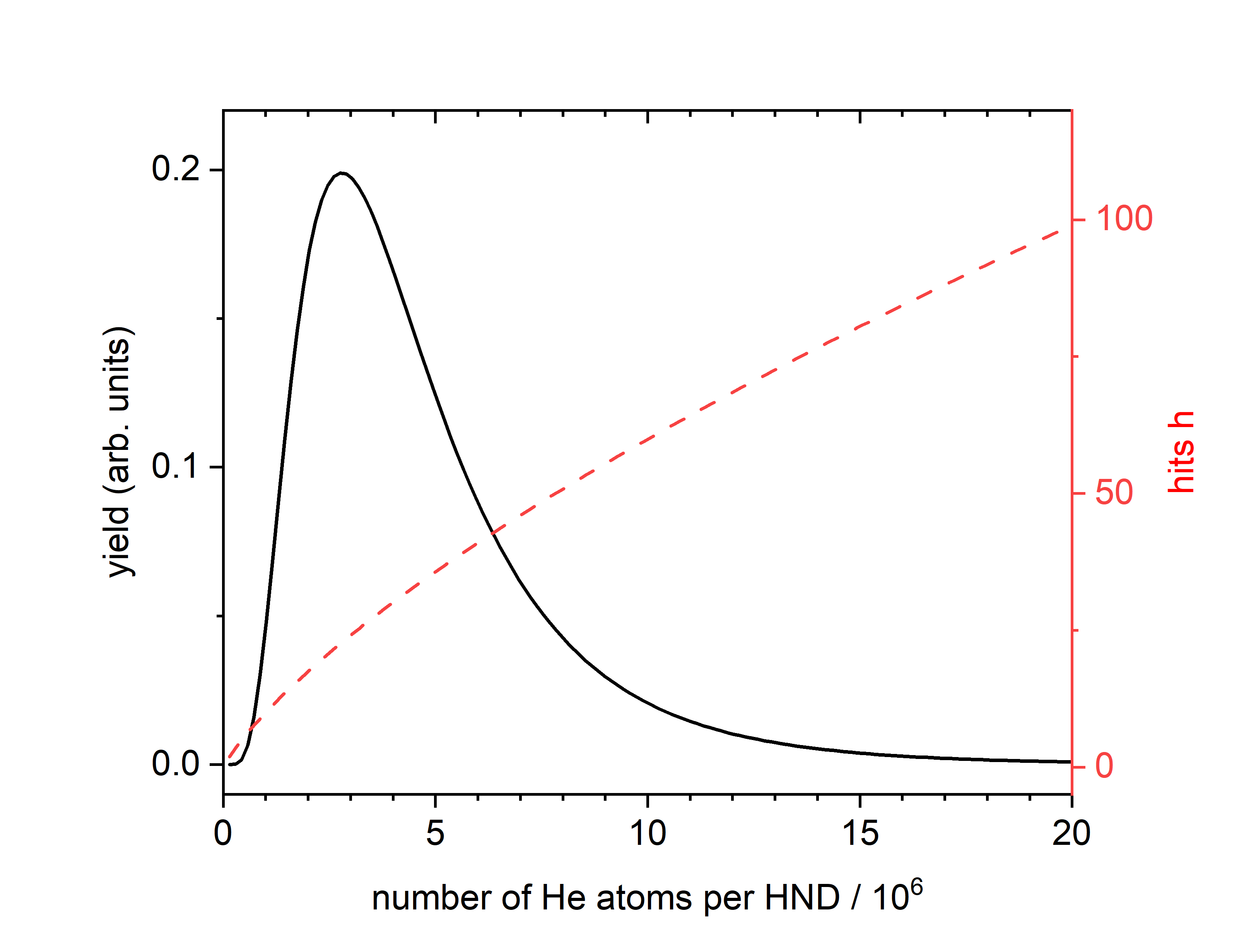}
    \caption{Size distribution of neutral HNDs formed at \SI{22}{bar} and \SI{8.5}{K}. The dashed red line shows the number of electron hits dependent on the HND size.}
    \label{SI_fig1}
\end{minipage}
\hfill
\begin{minipage}{0.48\textwidth}
\includegraphics[width=8cm]{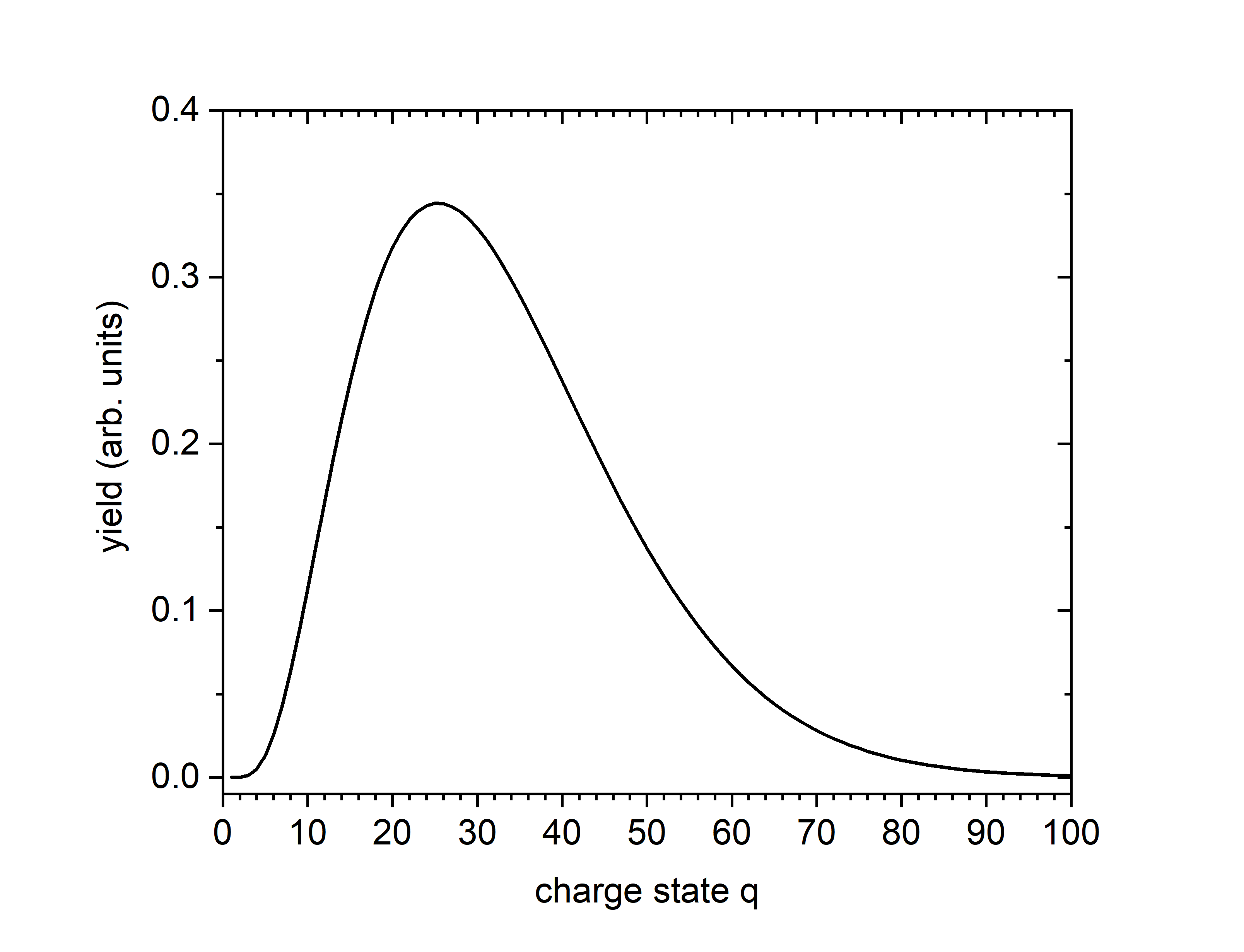}
    \caption{Calculated charge distribution of the size-per-charge selected HNDs (\SI{1.5e5}{} He atoms-per-charge) by using a sector field after the HND ionization.}
    \label{SI_fig2}
\end{minipage}
\end{figure}

With our HND source settings (helium stagnation pressure of 22~bar, 5~$\mu$m nozzle cooled to 8.5~K), we form HNDs that follow a log-normal distribution with a mean number of helium atoms of $\approx$\SI{4e6}{} per HND. Previous investigations of our group have revealed that the cluster size distribution can be described by the function $f(n)=\frac{1}{\sqrt{2\pi}\sigma n}\exp{(\frac{-[\ln (n/\mu)]^2}{2\sigma^2})}$, where $n$ is the number of helium atoms per HND and $\mu$ the mean number of helium atoms per HND. The fitting of the experimental data by using the function $f(n)$ revealed for $\mathrm{\sigma}$ the value 0.6 (see Fig.~\ref{SI_fig1}). 

Upon traversing the electron impact source, these droplets undergo multiple charging through electron impact ionization. Subsequent to the ionization, a sector field is used to selectively filter HNDs based on a specific kinetic energy which according to the low velocity spread of the HND beam is also proportional to the mass-to-charge ratio. For the presented data, we have set the sector voltages to choose HNDs that contain, on average, \SI{1.5e5}{} He atoms per charge.
Subsequently, we conduct calculations to determine both the charge distribution and the expected charge state for these selected ions. 

The electron beam, with an estimated diameter of \SI{1}{\milli\meter}, exhibits a flux of $\approx$\SI{300}{\mathrm{\mu}A/\milli\meter^{2}}. This corresponds to an electron density $\mathrm{\rho}_{e^{-}}$ of around \SI{1873}{e^{-}/(s~nm^{2})} as it impinges upon the HND beam. Given a HND velocity of $\approx$\SI{200}{\meter/\second}, the calculated interaction time $\mathrm{\tau}$ between the HND and the electron beam is $\approx$\SI{5}{\mathrm{\mu}\second}.

The number of electron hits \textit{h}, leading to the formation of He$^{+}$ through direct ionization or via Penning ionization involving two metastable He$^{*}$, is calculated as
$$h=\left(\sigma_{ion}+\frac{\sigma_{meta}}{2}\right)\times\rho_{e^{-}}\times t.$$

Here, $\sigma_{ion}$=\SI{1.7e-3}{\nano\meter}$^{2}$ and $\sigma_{meta}$=\SI{2.0e-4}{\nano\meter}$^{2}$ represent the cross-sections for the formation of He$^{+}$ and He$^{*}$ through electron impact at \SI{40}{eV}, respectively. By dividing $\sigma_{meta}$ by two, we 
account for the fact that two metastable He$^{*}$ are required for the formation of one He$^{+}$ via Penning ionization.

Considering the spherical geometry of HNDs and applying Beer Lambert's law for the electron passing through the HND, the corrected cross sections $\sigma_{ion_{corr}}$ and $\sigma_{meta_{corr}}$ are determined as:

\begin{figure}[h]
  \begin{minipage}{.5\textwidth}
    \begin{gather*}
        \sigma_{ion_{corr}}=\int_{0}^{R} 2\pi r\frac{\sigma_{ion}}{\sigma_{ion}+\sigma_{meta}}\left(1-e^{-2\sqrt{R^{2}-r^{2}}\lambda}\right)dr\\
        \sigma_{meta_{corr}}=\int_{0}^{R}  2\pi r\frac{\sigma_{meta}}{\sigma_{ion}+\sigma_{meta}}\left(1-e^{-2\sqrt{R^{2}-r^{2}}\lambda}\right)dr
    \end{gather*}
  \end{minipage}%
  \begin{minipage}{.5\textwidth}
    \centering
    \includegraphics[width=0.4\linewidth]{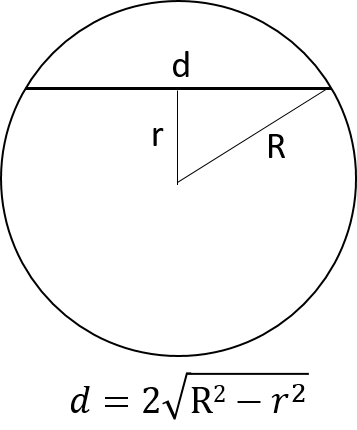}
  \end{minipage}
\end{figure}


Here, $\lambda = \rho_{He}(\sigma_{ion}+\sigma_{meta})$ represents the mean free path of electrons in liquid helium with a density of $\rho_{He}$=\SI{21.81}{He/nm^{3}}. By using this formula, we calculate the number of hits \textit{h} as a function of the HND radius \textit{R}, which can also be expressed as a function of the number of helium atoms per HND.

Assuming that the probability for the formation of \textit{q} charge centers in a HND follows the Poisson statistics for  certain number of hits $h$, we calculate $P_{h}(q)=\frac{h^q\times e^{-h}}{q!}$.
This formula is dependent on the number of helium atoms per HND, which is taken into account through the number of hits. By multiplying this function with the log-normal distribution of the initial cluster size (see Fig.~\ref{SI_fig1}), the probability of forming \textit{q} charge centers is weighted according to the initial HND size (see Fig.~\ref{SI_fig2}). The expected charge state value $\langle q \rangle$ is calculated to be 33 as a result of

$$\langle q \rangle = \frac{\sum q\times P_{h}(q)\times f(n)}{\sum P_{h}(q)\times f(n)}.$$

\section{Collision induced dissociation measurements}
\begin{figure}
\centering
\includegraphics[width=0.7\linewidth]{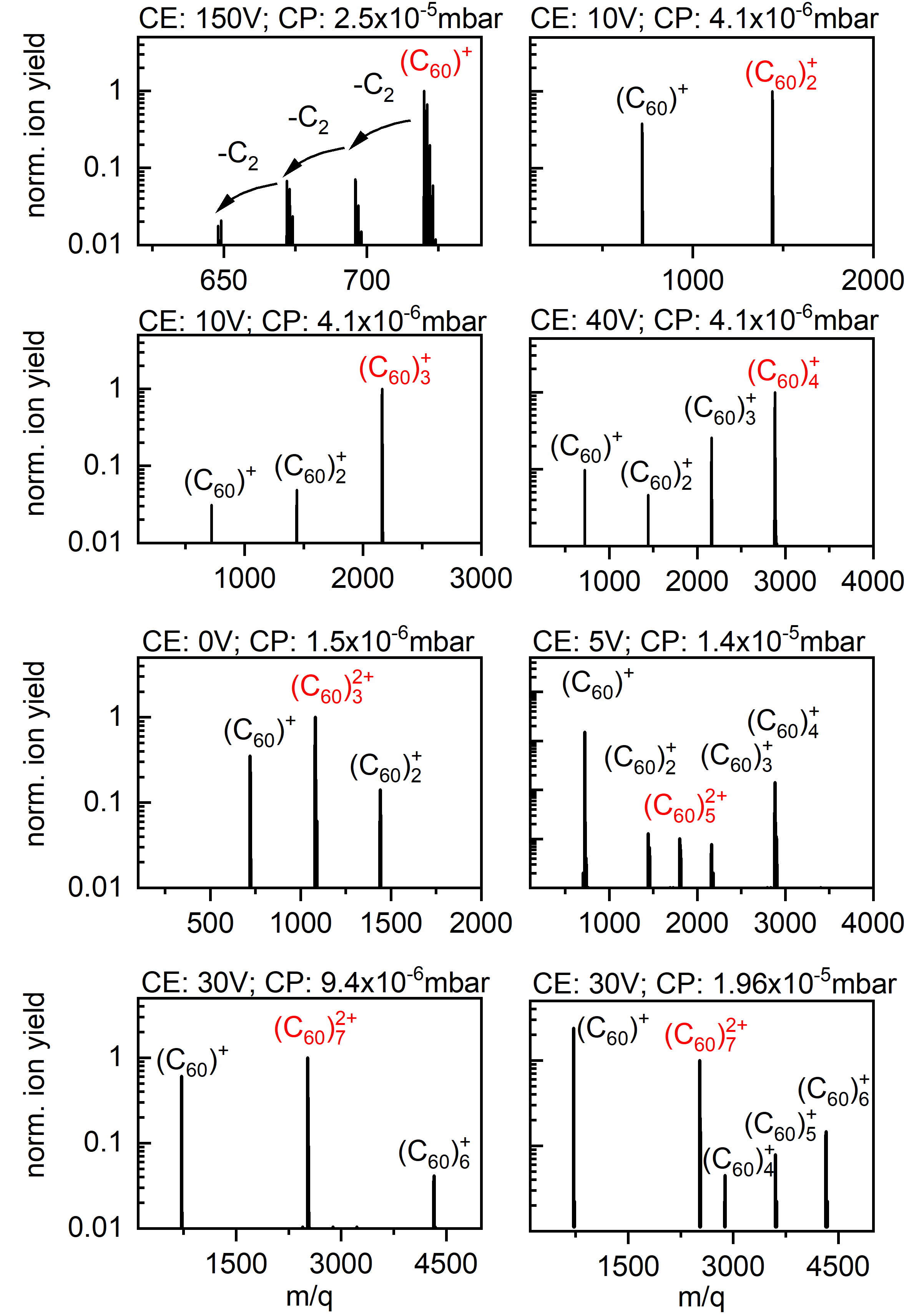}
\caption{Collision induced fragmentation studies of singly- and doubly-charged fullerene cluster ions across various collision energies CE and collision pressures CP. When considering the cation C$_{60}^{+}$, we observe the sequential loss of C$_{2}$ molecules. For C$_{60}$$_{n}^{+}$ with \textit{n} = 2, 3 and 4, we observe the sequential loss of entire C$_{60}$ fullerenes. In the case of doubly-charged cluster ions, C$_{60}$$_{n}^{2+}$, we note a preference for asymmetric splitting into singly-charged cluster ions. These singly-charged cluster ions subsequently dissociate further by losing C$_{60}$ fullerenes.} 
\label{SI_fig3} 
\end{figure}

After passing the second electron impact ionization source, the dopant ions are gently extracted from the HND by collisions with helium gas and afterwards filtered based on their specific mass-to-charge ratio by a RF-quadrupole. These selected ions are then accelerated into a collision chamber filled with argon gas at room temperature and various pressures, see Fig.~\ref{SI_fig3}, and the fragment ions are detected in the time-of-flight mass-spectrometer. When examining doubly-charged fullerene cluster ions, we focused on those with an odd number of C$_{60}$ units as they do not overlap with singly charged cluster ions.

\section{Quantum chemical calculations}

For neutral (C$_{60}$)$_2$, we started with the physisorbed structure with the two fullerene units oriented towards each other with hexagonal rings (IV), calculated to be slightly more stable (0.03 eV) than the one where two C$_{60}$ units face each other with pentagonal rings. From this structure, a selected C--C distance was scanned using a relaxed scan, i.e. optimizing all parameters with the exception of the C--C distance.

In (C$_{60}$)$_2^+$, we used relaxed scans along a C–-C bond, starting in the structure with two covalent bonds (II), forming eventually the structure with one covalent bond (III). Close to formation of the III structure, two fullerene units rotate against each other over the C–-C bond, this part of the potential energy surface was described using a relaxed scan over a C–-C–-C–-C dihedral angle along with a fixed C–-C distance. Subsequently, the second C–-C bond was scanned, reaching a physisorbed structure with two fullerene units facing each other through hexagons, IV. From this structure, a C–-C distance was further prolonged.

In (C$_{60}$)$_2^{2+}$ and (C$_{60}$)$_2^{3+}$, we used a similar approach as in (C$_{60}$)$_2^{+}$. Due to the converge issues with relaxed scans when starting with a physisorbed structure and prolonging the distance between the fullerene units, the Coulomb barrier was estimated by taking the structure optimized with the C–C distance of 3.8 Å and 3.2 Å for (C$_{60}$)$_2^{2+}$ and (C$_{60}$)$_2^{3+}$, respectively, prolonging it without re-optimization. This might lead to an overestimation of the Coulomb barrier. Singlet spin multiplicity was used for (C$_{60}$)$_2^{2+}$.

Note that at our level of theory (BP86-D3/def2SVP), ionization energies are underestimated by 0.3~eV and 0.9~eV for C$_{60}$ and C$_{60}^+$, respectively (4\% and 8\%, respectively).

\section{Cartesian coordinates of optimized structures (in \AA) along with the electronic energy (in Hartree)}

\obeylines

C$_{60}$
E = -2284.413714
C -0.000000 1.241985 3.349101
C 0.000000 2.440095 2.608629
C -1.181198 2.823889 1.841040
C -2.320668 0.754031 2.608629
C -1.181198 0.383795 3.349101
C -0.730021 -1.004787 3.349101
C 0.730021 -1.004787 3.349101
C 1.181198 0.383795 3.349101
C 0.000000 -1.241985 -3.349101
C -0.000000 -2.440095 -2.608629
C 1.181198 -2.823889 -1.841040
C 1.181198 -0.383795 -3.349101
C 2.320668 -1.996016 -1.841040
C 0.730021 1.004787 -3.349101
C -0.730021 1.004787 -3.349101
C -1.181198 -0.383795 -3.349101
C 1.181198 2.823889 1.841040
C -3.050689 1.758818 0.599055
C -2.320668 1.996016 1.841040
C -2.615450 -1.590284 1.841040
C 0.704231 -2.978865 1.841040
C 3.050689 -0.250756 1.841040
C -1.181198 -2.823889 -1.841040
C 3.050689 -1.758818 -0.599055
C 2.320668 1.996016 1.841040
C 2.615450 2.357873 -0.599055
C -1.434252 3.216063 -0.599055
C -3.501866 -0.370236 -0.599055
C -0.730021 -3.444882 -0.599055
C -2.615450 -2.357873 0.599055
C -0.704231 2.978865 -1.841040
C 1.434252 -3.216063 0.599055
C 3.501866 0.370236 0.599055
C 0.730021 3.444882 0.599055
C -3.050689 0.250756 -1.841040
C 2.615450 1.590284 -1.841040
C -0.730021 3.444882 0.599055
C -2.320668 -0.754031 -2.608629
C 2.320668 0.754031 2.608629
C -1.434252 -1.974078 2.608629
C 1.434252 -1.974078 2.608629
C 1.434252 3.216063 -0.599055
C -1.434252 1.974078 -2.608629
C -1.434252 -3.216063 0.599055
C 2.615450 -2.357873 0.599055
C 3.050689 1.758818 0.599055
C -3.501866 0.370236 0.599055
C -2.615450 2.357873 -0.599055
C -3.050689 -1.758818 -0.599055
C 0.730021 -3.444882 -0.599055
C 3.501866 -0.370236 -0.599055
C -2.615450 1.590284 -1.841040
C 1.434252 1.974078 -2.608629
C -2.320668 -1.996016 -1.841040
C 2.320668 -0.754031 -2.608629
C -3.050689 -0.250756 1.841040
C -0.704231 -2.978865 1.841040
C 2.615450 -1.590284 1.841040
C 3.050689 0.250756 -1.841040
C 0.704231 2.978865 -1.841040 \\

(C$_{60}$)$_2$, I
E = -4568.873128
C 7.425345 1.182048 1.434911
C 7.000746 2.328262 0.732603
C 7.000746 2.328262 -0.732603
C 7.425345 1.182048 -1.434911
C 7.873149 -0.000000 -0.704689
C 7.873149 -0.000000 0.704689
C 5.833569 3.071820 -1.184378
C 5.108371 2.628973 -2.313274
C 5.537536 1.432691 -3.024322
C 6.684749 0.730285 -2.607968
C 6.684749 -0.730285 -2.607968
C 5.537536 -1.432691 -3.024322
C 4.337050 -0.698362 -3.410989
C 4.337050 0.698362 -3.410989
C 3.170111 1.419340 -2.921775
C 2.009561 0.743545 -2.434700
C 2.009561 -0.743545 -2.434700
C 3.170111 -1.419340 -2.921775
C 5.108371 -2.628973 -2.313274
C 3.650997 -2.636061 -2.292618
C 3.650997 2.636061 -2.292618
C 2.969069 3.145396 -1.171532
C 1.744538 2.525208 -0.748527
C 1.216400 1.419879 -1.429163
C 0.000000 0.773687 -0.813767
C 1.216400 -1.419879 -1.429163
C 1.744538 -2.525208 -0.748527
C 2.969069 -3.145396 -1.171532
C 7.425345 -1.182048 -1.434911
C 7.000746 -2.328262 -0.732603
C 5.833569 -3.071820 -1.184378
C 5.120237 -3.545325 0.000000
C 3.718106 -3.580448 0.000000
C -0.000000 0.773687 0.813767
C 1.216400 1.419879 1.429163
C 1.744538 2.525208 0.748527
C 2.969069 3.145396 1.171532
C 3.718106 3.580448 0.000000
C 5.120237 3.545325 0.000000
C 5.833569 3.071820 1.184378
C 5.108371 2.628973 2.313274
C 3.650997 2.636061 2.292618
C 2.009561 0.743545 2.434700
C 3.170111 1.419340 2.921775
C 6.684749 0.730285 2.607968
C 5.537536 1.432691 3.024322
C 4.337050 0.698362 3.410989
C 1.216400 -1.419879 1.429163
C 1.744538 -2.525208 0.748527
C 2.969069 -3.145396 1.171532
C 7.425345 -1.182048 1.434911
C 7.000746 -2.328262 0.732603
C 5.833569 -3.071820 1.184378
C 5.108371 -2.628973 2.313274
C 3.650997 -2.636061 2.292618
C 2.009561 -0.743545 2.434700
C 3.170111 -1.419340 2.921775
C 4.337050 -0.698362 3.410989
C 5.537536 -1.432691 3.024322
C 6.684749 -0.730285 2.607968
C -1.216400 -1.419879 -1.429163
C -1.744538 -2.525208 -0.748527
C -1.744538 -2.525208 0.748527
C -1.216400 -1.419879 1.429163
C -0.000000 -0.773687 0.813767
C -0.000000 -0.773687 -0.813767
C -2.009561 -0.743545 2.434700
C -3.170111 -1.419340 2.921775
C -3.650997 -2.636061 2.292618
C -2.969069 -3.145396 1.171532
C -4.337050 -0.698362 3.410989
C -4.337050 0.698362 3.410989
C -3.170111 1.419340 2.921775
C -2.009561 0.743545 2.434700
C -1.216400 1.419879 1.429163
C -1.744538 2.525208 0.748527
C -2.969069 3.145396 1.171532
C -3.650997 2.636061 2.292618
C -1.744538 2.525208 -0.748527
C -2.969069 3.145396 -1.171532
C -3.718106 3.580448 0.000000
C -5.120237 3.545325 0.000000
C -5.833569 3.071820 -1.184378
C -5.108371 2.628973 -2.313274
C -3.650997 2.636061 -2.292618
C -3.170111 1.419340 -2.921775
C -2.009561 0.743545 -2.434700
C -1.216400 1.419879 -1.429163
C -5.833569 3.071820 1.184378
C -5.108371 2.628973 2.313274
C -5.537536 1.432691 3.024322
C -5.108371 -2.628973 2.313274
C -5.537536 -1.432691 3.024322
C -6.684749 -0.730285 2.607968
C -6.684749 0.730285 2.607968
C -5.833569 -3.071820 1.184378
C -5.120237 -3.545325 -0.000000
C -3.718106 -3.580448 -0.000000
C -2.969069 -3.145396 -1.171532
C -7.000746 -2.328262 0.732603
C -7.425345 -1.182048 1.434911
C -7.000746 2.328262 0.732603
C -7.425345 1.182048 1.434911
C -7.873149 -0.000000 0.704689
C -7.873149 0.000000 -0.704689
C -7.425345 -1.182048 -1.434911
C -7.000746 -2.328262 -0.732603
C -5.833569 -3.071820 -1.184378
C -5.108371 -2.628973 -2.313274
C -3.650997 -2.636061 -2.292618
C -7.425345 1.182048 -1.434911
C -7.000746 2.328262 -0.732603
C -6.684749 0.730285 -2.607968
C -6.684749 -0.730285 -2.607968
C -5.537536 -1.432691 -3.024322
C -4.337050 -0.698362 -3.410989
C -3.170111 -1.419340 -2.921775
C -2.009561 -0.743545 -2.434700
C -4.337050 0.698362 -3.410989
C -5.537536 1.432691 -3.024322 \\

(C$_{60}$)$_2$, II
E = -4568.851271
C -7.603222 1.435829 1.180765
C -7.163085 0.731442 2.319397
C -7.163085 -0.731442 2.319397
C -7.603222 -1.435829 1.180765
C -8.051949 -0.704395 0.000000
C -8.051949 0.704395 0.000000
C -5.984135 -1.182853 3.045513
C -5.283668 -2.327558 2.614567
C -5.740747 -3.059482 1.435139
C -6.878844 -2.619559 0.730507
C -6.878844 -2.619559 -0.730507
C -5.740747 -3.059482 -1.435139
C -4.562400 -3.513405 -0.703822
C -4.562400 -3.513405 0.703822
C -3.382274 -3.059639 1.435936
C -2.247822 -2.605353 0.729671
C -2.247822 -2.605353 -0.729671
C -3.382274 -3.059639 -1.435936
C -5.283668 -2.327558 -2.614567
C -3.825583 -2.330985 -2.609906
C -3.825583 -2.330985 2.609906
C -3.114054 -1.182563 3.023043
C -1.934281 -0.742925 2.294915
C -1.497977 -1.456437 1.181841
C -0.795186 -0.800313 0.000000
C -1.497977 -1.456437 -1.181841
C -1.934281 -0.742925 -2.294915
C -3.114054 -1.182563 -3.023043
C -7.603222 -1.435829 -1.180765
C -7.163085 -0.731442 -2.319397
C -5.984135 -1.182853 -3.045513
C -5.249669 0.000000 -3.488397
C -3.842457 0.000000 -3.470058
C -0.795186 0.800313 0.000000
C -1.497977 1.456437 1.181841
C -1.934281 0.742925 2.294915
C -3.114054 1.182563 3.023043
C -3.842457 -0.000000 3.470058
C -5.249669 -0.000000 3.488397
C -5.984135 1.182853 3.045513
C -5.283668 2.327558 2.614567
C -3.825583 2.330985 2.609906
C -2.247822 2.605353 0.729671
C -3.382274 3.059639 1.435936
C -6.878844 2.619559 0.730507
C -5.740747 3.059482 1.435139
C -4.562400 3.513405 0.703822
C -1.497977 1.456437 -1.181841
C -1.934281 0.742925 -2.294915
C -3.114054 1.182563 -3.023043
C -7.603222 1.435829 -1.180765
C -7.163085 0.731442 -2.319397
C -5.984135 1.182853 -3.045513
C -5.283668 2.327558 -2.614567
C -3.825583 2.330985 -2.609906
C -2.247822 2.605353 -0.729671
C -3.382274 3.059639 -1.435936
C -4.562400 3.513405 -0.703822
C -5.740747 3.059482 -1.435139
C -6.878844 2.619559 -0.730507
C 1.497977 -1.456437 -1.181841
C 1.934281 -0.742925 -2.294915
C 1.934281 0.742925 -2.294915
C 1.497977 1.456437 -1.181841
C 0.795186 0.800313 0.000000
C 0.795186 -0.800313 0.000000
C 2.247822 2.605353 -0.729671
C 3.382274 3.059639 -1.435936
C 3.825583 2.330985 -2.609906
C 3.114054 1.182563 -3.023043
C 4.562400 3.513405 -0.703822
C 4.562400 3.513405 0.703822
C 3.382274 3.059639 1.435936
C 2.247822 2.605353 0.729671
C 1.497977 1.456437 1.181841
C 1.934281 0.742925 2.294915
C 3.114054 1.182563 3.023043
C 3.825583 2.330985 2.609906
C 1.934281 -0.742925 2.294915
C 3.114054 -1.182563 3.023043
C 3.842457 0.000000 3.470058
C 5.249669 0.000000 3.488397
C 5.984135 -1.182853 3.045513
C 5.283668 -2.327558 2.614567
C 3.825583 -2.330985 2.609906
C 3.382274 -3.059639 1.435936
C 2.247822 -2.605353 0.729671
C 1.497977 -1.456437 1.181841
C 5.984135 1.182853 3.045513
C 5.283668 2.327558 2.614567
C 5.740747 3.059482 1.435139
C 5.283668 2.327558 -2.614567
C 5.740747 3.059482 -1.435139
C 6.878844 2.619559 -0.730507
C 6.878844 2.619559 0.730507
C 5.984135 1.182853 -3.045513
C 5.249669 -0.000000 -3.488397
C 3.842457 -0.000000 -3.470058
C 3.114054 -1.182563 -3.023043
C 7.163085 0.731442 -2.319397
C 7.603222 1.435829 -1.180765
C 7.163085 0.731442 2.319397
C 7.603222 1.435829 1.180765
C 8.051949 0.704395 0.000000
C 8.051949 -0.704395 0.000000
C 7.603222 -1.435829 -1.180765
C 7.163085 -0.731442 -2.319397
C 5.984135 -1.182853 -3.045513
C 5.283668 -2.327558 -2.614567
C 3.825583 -2.330985 -2.609906
C 7.603222 -1.435829 1.180765
C 7.163085 -0.731442 2.319397
C 6.878844 -2.619559 0.730507
C 6.878844 -2.619559 -0.730507
C 5.740747 -3.059482 -1.435139
C 4.562400 -3.513405 -0.703822
C 3.382274 -3.059639 -1.435936
C 2.247822 -2.605353 -0.729671
C 4.562400 -3.513405 0.703822
C 5.740747 -3.059482 1.435139 \\

(C$_{60}$)$_2$, IV
E = -4568.842716
c 2.829459 -1.377927 -2.574854
c 7.366803 -0.403858 -2.538928
c 6.409596 -1.359563 -2.931647
c 6.198753 -2.556892 -2.122529
c 6.952688 -2.755510 -0.949534
c 7.945348 -1.765093 -0.542816
c 8.148937 -0.610379 -1.323271
c 5.112921 -0.925985 -3.442868
c 4.100098 -1.855400 -2.949810
c 4.771040 -2.863025 -2.133351
c 4.819153 0.447924 -3.543640
c 3.501378 0.943740 -3.154920
c 2.524755 0.047163 -2.679018
c 2.184466 -1.891366 -1.369120
c 1.485644 -0.782931 -0.727724
c 1.694037 0.414842 -1.537061
c 3.679465 2.240370 -2.508327
c 2.874540 2.595315 -1.407507
c 1.863294 1.666006 -0.913690
c 1.830360 1.765804 0.542794
c 1.632598 0.609744 1.322285
c 1.456302 -0.686845 0.675735
c 2.120307 -1.695561 1.492882
c 2.798134 -2.764712 0.874960
c 2.830657 -2.864284 -0.582006
c 4.147849 -3.358171 -0.970888
c 5.107168 2.545970 -2.496552
c 5.811594 1.438548 -3.136882
c 7.062408 1.020115 -2.643379
c 7.656024 1.694140 -1.492108
c 6.976191 2.762075 -0.874529
c 5.679320 3.195967 -1.385803
c 3.467809 3.269164 -0.255659
c 4.845093 3.563265 -0.245227
c 2.822041 2.756912 0.949432
c 5.626879 3.357075 0.970765
c 5.003659 2.863131 2.133218
c 3.575838 2.558144 2.123251
c 2.408795 0.403832 2.540396
c 3.365554 1.360483 2.932689
c 2.712840 -1.020765 2.644888
c 4.662495 0.926567 3.443438
c 4.956146 -0.447611 3.543451
c 3.963576 -1.439143 3.137955
c 4.095437 -3.197463 1.386037
c 4.667658 -2.546398 2.496631
c 4.929680 -3.563790 0.245109
c 6.306776 -3.267994 0.255346
c 6.095005 -2.239704 2.507172
c 6.900204 -2.594520 1.406971
c 7.912677 -1.665349 0.913415
c 8.084234 -0.414391 1.537615
c 7.250074 -0.046900 2.677967
c 6.273481 -0.942965 3.153986
c 8.295440 0.782562 0.728645
c 7.591111 1.889826 1.368652
c 6.945452 1.377506 2.573973
c 5.674975 1.855183 2.949317
c 6.943649 2.861907 0.581811
c 8.327262 0.686381 -0.676186
c -2.839916 0.261954 -2.916163
c -6.947762 2.184968 -1.932909
c -6.201871 1.543173 -2.940546
c -6.422710 0.127794 -3.222617
c -7.381510 -0.595850 -2.487128
c -8.154981 0.069438 -1.442576
c -7.942052 1.435107 -1.170575
c -4.773546 1.813305 -3.075781
c -4.111255 0.564714 -3.441235
c -5.130551 -0.477331 -3.531381
c -4.141142 2.716256 -2.198238
c -2.823096 2.403154 -1.653644
c -2.184921 1.198261 -2.006222
c -2.541487 -1.094123 -2.462399
c -1.704591 -0.996116 -1.270665
c -1.486795 0.420350 -0.988739
c -2.781857 2.875205 -0.271903
c -2.104816 2.124772 0.708832
c -1.449898 0.874485 0.342562
c -1.627019 -0.069371 1.441172
c -1.833739 -1.435884 1.170534
c -1.875626 -1.907025 -0.210850
c -2.893944 -2.948513 -0.301758
c -3.703936 -3.042623 -1.450596
c -3.524549 -2.099078 -2.549997
c -4.842531 -1.784796 -3.094431
c -4.074168 3.479700 0.037195
c -4.914084 3.381069 -1.153144
c -6.292107 3.120062 -1.023111
c -6.880716 2.947365 0.301491
c -6.070565 3.041544 1.449734
c -4.642288 3.312978 1.315858
c -2.692289 1.952596 2.035224
c -3.938913 2.536459 2.333095
c -2.394141 0.596407 2.488502
c -4.932687 1.784556 3.094366
c -4.644691 0.477119 3.532124
c -3.352333 -0.128233 3.223929
c -2.826873 -2.186388 1.933502
c -3.572521 -1.544139 2.941884
c -3.482469 -3.121457 1.023389
c -5.001080 -1.813351 3.075814
c -5.633567 -2.715274 2.197825
c -4.860799 -3.380663 1.152992
c -5.132630 -3.312448 -1.315871
c -5.700703 -3.478092 -0.037423
c -5.836249 -2.535368 -2.332290
c -7.083042 -1.951321 -2.033993
c -6.992563 -2.872462 0.271408
c -7.671763 -2.123254 -0.708652
c -8.334257 -0.874461 -0.343273
c -8.294244 -0.419656 0.989445
c -7.590542 -1.196857 2.005264
c -6.951147 -2.400908 1.652734
c -8.073373 0.995864 1.271115
c -7.233254 1.093507 2.461359
c -6.934953 -0.261798 2.915190
c -5.663698 -0.564596 3.440714
c -6.250275 2.098017 2.549327
c -7.900043 1.906088 0.210576 \\

C$_{60}^+$
E = -2284.144595
C 0.723156 3.425701 0.603932
C 3.043103 0.250796 -1.855458
C 2.609337 1.585793 -1.855458
C 1.433662 1.973267 -2.627462
C 0.730111 1.004911 -3.368743
C 1.181344 -0.383842 -3.368743
C 2.319714 -0.753721 -2.627462
C 2.598622 2.346390 -0.603932
C 1.428531 3.196510 -0.603932
C 0.701849 2.971664 -1.855458
C 3.034568 1.746362 0.603932
C 2.314508 1.991590 1.855458
C 1.178892 2.816663 1.855458
C -0.723156 3.425701 0.603932
C -1.178892 2.816663 1.855458
C 0.000000 2.439092 2.627462
C 2.319714 0.753721 2.627462
C 1.181344 0.383842 3.368743
C 0.000000 1.242138 3.368743
C -1.181344 0.383842 3.368743
C -2.319714 0.753721 2.627462
C -2.314508 1.991590 1.855458
C -3.034568 1.746362 0.603932
C -2.598622 2.346390 -0.603932
C -1.428531 3.196510 -0.603932
C -0.701849 2.971664 -1.855458
C 3.043103 -0.250796 1.855458
C 3.481503 0.370838 0.603932
C 3.481503 -0.370838 -0.603932
C 3.034568 -1.746362 -0.603932
C 2.598622 -2.346390 0.603932
C 2.609337 -1.585793 1.855458
C 0.730111 -1.004911 3.368743
C 1.433662 -1.973267 2.627462
C -0.730111 -1.004911 3.368743
C 0.701849 -2.971664 1.855458
C -0.701849 -2.971664 1.855458
C -1.433662 -1.973267 2.627462
C -3.043103 -0.250796 1.855458
C -2.609337 -1.585793 1.855458
C -3.481503 0.370838 0.603932
C -2.598622 -2.346390 0.603932
C -3.034568 -1.746362 -0.603932
C -3.481503 -0.370838 -0.603932
C -2.609337 1.585793 -1.855458
C -3.043103 0.250796 -1.855458
C -1.433662 1.973267 -2.627462
C -0.730111 1.004911 -3.368743
C -2.319714 -0.753721 -2.627462
C -1.181344 -0.383842 -3.368743
C 0.000000 -1.242138 -3.368743
C 0.000000 -2.439092 -2.627462
C -1.178892 -2.816663 -1.855458
C -2.314508 -1.991590 -1.855458
C 1.178892 -2.816663 -1.855458
C 0.723156 -3.425701 -0.603932
C -0.723156 -3.425701 -0.603932
C -1.428531 -3.196510 0.603932
C 1.428531 -3.196510 0.603932
C 2.314508 -1.991590 -1.855458 \\

(C$_{60}$)$_2^+$, I
E = -4568.629627
C 7.422218 1.182065 1.433352
C 6.998920 2.329074 0.731084
C 6.998920 2.329074 -0.731084
C 7.422218 1.182065 -1.433352
C 7.870927 -0.000000 -0.704194
C 7.870927 -0.000000 0.704194
C 5.831853 3.076033 -1.183306
C 5.107890 2.634235 -2.311074
C 5.534439 1.436726 -3.020070
C 6.678304 0.731420 -2.602952
C 6.678304 -0.731420 -2.602952
C 5.534439 -1.436726 -3.020070
C 4.332079 -0.701123 -3.395871
C 4.332079 0.701123 -3.395871
C 3.168330 1.419924 -2.912362
C 2.005612 0.744037 -2.428225
C 2.005612 -0.744037 -2.428225
C 3.168330 -1.419924 -2.912362
C 5.107890 -2.634235 -2.311074
C 3.649569 -2.641071 -2.286393
C 3.649569 2.641071 -2.286393
C 2.969048 3.154831 -1.170035
C 1.743847 2.528929 -0.744175
C 1.212946 1.420255 -1.427861
C 0.000000 0.773182 -0.811833
C 1.212946 -1.420255 -1.427861
C 1.743847 -2.528929 -0.744175
C 2.969048 -3.154831 -1.170035
C 7.422218 -1.182065 -1.433352
C 6.998920 -2.329074 -0.731084
C 5.831853 -3.076033 -1.183306
C 5.119358 -3.552609 0.000000
C 3.717517 -3.593566 0.000000
C -0.000000 0.773182 0.811833
C 1.212946 1.420255 1.427861
C 1.743847 2.528929 0.744175
C 2.969048 3.154831 1.170035
C 3.717517 3.593566 0.000000
C 5.119358 3.552609 0.000000
C 5.831853 3.076033 1.183306
C 5.107890 2.634235 2.311074
C 3.649569 2.641071 2.286393
C 2.005612 0.744037 2.428225
C 3.168330 1.419924 2.912362
C 6.678304 0.731420 2.602952
C 5.534439 1.436726 3.020070
C 4.332079 0.701123 3.395871
C 1.212946 -1.420255 1.427861
C 1.743847 -2.528929 0.744175
C 2.969048 -3.154831 1.170035
C 7.422218 -1.182065 1.433352
C 6.998920 -2.329074 0.731084
C 5.831853 -3.076033 1.183306
C 5.107890 -2.634235 2.311074
C 3.649569 -2.641071 2.286393
C 2.005612 -0.744037 2.428225
C 3.168330 -1.419924 2.912362
C 4.332079 -0.701123 3.395871
C 5.534439 -1.436726 3.020070
C 6.678304 -0.731420 2.602952
C -1.212946 -1.420255 -1.427861
C -1.743847 -2.528929 -0.744175
C -1.743847 -2.528929 0.744175
C -1.212946 -1.420255 1.427861
C -0.000000 -0.773182 0.811833
C -0.000000 -0.773182 -0.811833
C -2.005612 -0.744037 2.428225
C -3.168330 -1.419924 2.912362
C -3.649569 -2.641071 2.286393
C -2.969048 -3.154831 1.170035
C -4.332079 -0.701123 3.395871
C -4.332079 0.701123 3.395871
C -3.168330 1.419924 2.912362
C -2.005612 0.744037 2.428225
C -1.212946 1.420255 1.427861
C -1.743847 2.528929 0.744175
C -2.969048 3.154831 1.170035
C -3.649569 2.641071 2.286393
C -1.743847 2.528929 -0.744175
C -2.969048 3.154831 -1.170035
C -3.717517 3.593566 0.000000
C -5.119358 3.552609 0.000000
C -5.831853 3.076033 -1.183306
C -5.107890 2.634235 -2.311074
C -3.649569 2.641071 -2.286393
C -3.168330 1.419924 -2.912362
C -2.005612 0.744037 -2.428225
C -1.212946 1.420255 -1.427861
C -5.831853 3.076033 1.183306
C -5.107890 2.634235 2.311074
C -5.534439 1.436726 3.020070
C -5.107890 -2.634235 2.311074
C -5.534439 -1.436726 3.020070
C -6.678304 -0.731420 2.602952
C -6.678304 0.731420 2.602952
C -5.831853 -3.076033 1.183306
C -5.119358 -3.552609 -0.000000
C -3.717517 -3.593566 -0.000000
C -2.969048 -3.154831 -1.170035
C -6.998920 -2.329074 0.731084
C -7.422218 -1.182065 1.433352
C -6.998920 2.329074 0.731084
C -7.422218 1.182065 1.433352
C -7.870927 -0.000000 0.704194
C -7.870927 0.000000 -0.704194
C -7.422218 -1.182065 -1.433352
C -6.998920 -2.329074 -0.731084
C -5.831853 -3.076033 -1.183306
C -5.107890 -2.634235 -2.311074
C -3.649569 -2.641071 -2.286393
C -7.422218 1.182065 -1.433352
C -6.998920 2.329074 -0.731084
C -6.678304 0.731420 -2.602952
C -6.678304 -0.731420 -2.602952
C -5.534439 -1.436726 -3.020070
C -4.332079 -0.701123 -3.395871
C -3.168330 -1.419924 -2.912362
C -2.005612 -0.744037 -2.428225
C -4.332079 0.701123 -3.395871
C -5.534439 1.436726 -3.020070 \\

(C$_{60}$)$_2^+$, II
E = -4568.601174
C -7.598165 1.433985 1.181432
C -7.158559 0.729828 2.321577
C -7.158559 -0.729828 2.321577
C -7.598165 -1.433985 1.181432
C -8.048010 -0.703353 0.000000
C -8.048010 0.703353 0.000000
C -5.980266 -1.182142 3.051269
C -5.281074 -2.326085 2.621530
C -5.736124 -3.054429 1.439844
C -6.871105 -2.612709 0.732298
C -6.871105 -2.612709 -0.732298
C -5.736124 -3.054429 -1.439844
C -4.556798 -3.497760 -0.707311
C -4.556798 -3.497760 0.707311
C -3.379012 -3.052578 1.438639
C -2.243322 -2.599604 0.731121
C -2.243322 -2.599604 -0.731121
C -3.379012 -3.052578 -1.438639
C -5.281074 -2.326085 -2.621530
C -3.822201 -2.327412 -2.616318
C -3.822201 -2.327412 2.616318
C -3.111970 -1.181625 3.031678
C -1.933063 -0.740050 2.298638
C -1.494212 -1.455730 1.183238
C -0.792687 -0.798811 0.000000
C -1.494212 -1.455730 -1.183238
C -1.933063 -0.740050 -2.298638
C -3.111970 -1.181625 -3.031678
C -7.598165 -1.433985 -1.181432
C -7.158559 -0.729828 -2.321577
C -5.980266 -1.182142 -3.051269
C -5.246474 0.000000 -3.496398
C -3.839558 0.000000 -3.481322
C -0.792687 0.798811 0.000000
C -1.494212 1.455730 1.183238
C -1.933063 0.740050 2.298638
C -3.111970 1.181625 3.031678
C -3.839558 -0.000000 3.481322
C -5.246474 -0.000000 3.496398
C -5.980266 1.182142 3.051269
C -5.281074 2.326085 2.621530
C -3.822201 2.327412 2.616318
C -2.243322 2.599604 0.731121
C -3.379012 3.052578 1.438639
C -6.871105 2.612709 0.732298
C -5.736124 3.054429 1.439844
C -4.556798 3.497760 0.707311
C -1.494212 1.455730 -1.183238
C -1.933063 0.740050 -2.298638
C -3.111970 1.181625 -3.031678
C -7.598165 1.433985 -1.181432
C -7.158559 0.729828 -2.321577
C -5.980266 1.182142 -3.051269
C -5.281074 2.326085 -2.621530
C -3.822201 2.327412 -2.616318
C -2.243322 2.599604 -0.731121
C -3.379012 3.052578 -1.438639
C -4.556798 3.497760 -0.707311
C -5.736124 3.054429 -1.439844
C -6.871105 2.612709 -0.732298
C 1.494212 -1.455730 -1.183238
C 1.933063 -0.740050 -2.298638
C 1.933063 0.740050 -2.298638
C 1.494212 1.455730 -1.183238
C 0.792687 0.798811 0.000000
C 0.792687 -0.798811 0.000000
C 2.243322 2.599604 -0.731121
C 3.379012 3.052578 -1.438639
C 3.822201 2.327412 -2.616318
C 3.111970 1.181625 -3.031678
C 4.556798 3.497760 -0.707311
C 4.556798 3.497760 0.707311
C 3.379012 3.052578 1.438639
C 2.243322 2.599604 0.731121
C 1.494212 1.455730 1.183238
C 1.933063 0.740050 2.298638
C 3.111970 1.181625 3.031678
C 3.822201 2.327412 2.616318
C 1.933063 -0.740050 2.298638
C 3.111970 -1.181625 3.031678
C 3.839558 0.000000 3.481322
C 5.246474 0.000000 3.496398
C 5.980266 -1.182142 3.051269
C 5.281074 -2.326085 2.621530
C 3.822201 -2.327412 2.616318
C 3.379012 -3.052578 1.438639
C 2.243322 -2.599604 0.731121
C 1.494212 -1.455730 1.183238
C 5.980266 1.182142 3.051269
C 5.281074 2.326085 2.621530
C 5.736124 3.054429 1.439844
C 5.281074 2.326085 -2.621530
C 5.736124 3.054429 -1.439844
C 6.871105 2.612709 -0.732298
C 6.871105 2.612709 0.732298
C 5.980266 1.182142 -3.051269
C 5.246474 -0.000000 -3.496398
C 3.839558 -0.000000 -3.481322
C 3.111970 -1.181625 -3.031678
C 7.158559 0.729828 -2.321577
C 7.598165 1.433985 -1.181432
C 7.158559 0.729828 2.321577
C 7.598165 1.433985 1.181432
C 8.048010 0.703353 0.000000
C 8.048010 -0.703353 0.000000
C 7.598165 -1.433985 -1.181432
C 7.158559 -0.729828 -2.321577
C 5.980266 -1.182142 -3.051269
C 5.281074 -2.326085 -2.621530
C 3.822201 -2.327412 -2.616318
C 7.598165 -1.433985 1.181432
C 7.158559 -0.729828 2.321577
C 6.871105 -2.612709 0.732298
C 6.871105 -2.612709 -0.732298
C 5.736124 -3.054429 -1.439844
C 4.556798 -3.497760 -0.707311
C 3.379012 -3.052578 -1.438639
C 2.243322 -2.599604 -0.731121
C 4.556798 -3.497760 0.707311
C 5.736124 -3.054429 1.439844 \\

(C$_{60}$)$_2^+$, III
E = -4568.583742
C 0.011079 0.845087 0.000000
C -0.011079 -0.845087 0.000000
C -3.230979 4.083132 1.431331
C -3.035920 2.882912 0.727463
C -3.035920 2.882912 -0.727463
C -3.230979 4.083132 -1.431331
C -3.442155 5.330242 -0.704556
C -3.442155 5.330242 0.704556
C -2.003673 1.940675 1.168391
C -1.384039 1.380548 0.000000
C -2.003673 1.940675 -1.168391
C -1.225859 2.226299 2.315027
C 0.213569 1.931179 2.313534
C 0.817698 1.366769 1.188361
C 0.817698 1.366769 -1.188361
C 2.092785 1.855469 -0.731436
C 2.092785 1.855469 0.731436
C 0.882787 2.998987 3.043959
C 2.142516 3.463937 2.618834
C 2.762334 2.881169 1.438915
C 3.432695 3.946996 0.705819
C 3.432695 3.946996 -0.705819
C 2.762334 2.881169 -1.438915
C 2.142516 3.463937 -2.618834
C 0.882787 2.998987 -3.043959
C 0.213569 1.931179 -2.313534
C -1.225859 2.226299 -2.315027
C -0.134561 3.944593 3.497975
C -1.438019 3.460777 3.048724
C -2.417978 4.379989 2.610833
C -2.133711 5.806746 2.613703
C -0.874087 6.271116 3.045988
C 0.142236 5.322435 3.499047
C 2.427860 4.895773 2.618744
C 1.444951 5.807324 3.050453
C 3.230264 5.194421 1.436454
C 1.231213 7.048729 2.317774
C 2.009203 7.340788 1.179353
C 3.024781 6.394859 0.730312
C 3.230264 5.194421 -1.436454
C 3.024781 6.394859 -0.730312
C 2.427860 4.895773 -2.618744
C 2.009203 7.340788 -1.179353
C 1.231213 7.048729 -2.317774
C 1.444951 5.807324 -3.050453
C -0.134561 3.944593 -3.497975
C 0.142236 5.322435 -3.499047
C -1.438019 3.460777 -3.048724
C -2.417978 4.379989 -2.610833
C -0.874087 6.271116 -3.045988
C -2.133711 5.806746 -2.613703
C -2.763941 6.396911 -1.435029
C -2.110184 7.424058 -0.730399
C -0.806371 7.904647 -1.180023
C -0.201764 7.338548 -2.319023
C -2.110184 7.424058 0.730399
C -0.806371 7.904647 1.180023
C -0.003066 8.200416 0.000000
C 1.380584 7.924119 0.000000
C -0.201764 7.338548 2.319023
C -2.763941 6.396911 1.435029
C 2.003673 -1.940675 1.168391
C 3.035920 -2.882912 0.727463
C 3.035920 -2.882912 -0.727463
C 2.003673 -1.940675 -1.168391
C 1.384039 -1.380548 0.000000
C 3.230979 -4.083132 1.431331
C 2.417978 -4.379989 2.610833
C 1.438019 -3.460777 3.048724
C 1.225859 -2.226299 2.315027
C -0.213569 -1.931179 2.313534
C -0.817698 -1.366769 1.188361
C 3.442155 -5.330242 0.704556
C 3.442155 -5.330242 -0.704556
C 3.230979 -4.083132 -1.431331
C 2.417978 -4.379989 -2.610833
C 1.438019 -3.460777 -3.048724
C 1.225859 -2.226299 -2.315027
C -0.817698 -1.366769 -1.188361
C -0.213569 -1.931179 -2.313534
C 0.134561 -3.944593 3.497975
C -0.882787 -2.998987 3.043959
C -2.092785 -1.855469 0.731436
C -2.762334 -2.881169 1.438915
C -2.142516 -3.463937 2.618834
C -2.427860 -4.895773 2.618744
C -1.444951 -5.807324 3.050453
C -0.142236 -5.322435 3.499047
C 2.133711 -5.806746 2.613703
C 0.874087 -6.271116 3.045988
C 2.763941 -6.396911 1.435029
C 0.201764 -7.338548 2.319023
C 0.806371 -7.904647 1.180023
C 2.110184 -7.424058 0.730399
C 2.763941 -6.396911 -1.435029
C 2.110184 -7.424058 -0.730399
C 2.133711 -5.806746 -2.613703
C 0.806371 -7.904647 -1.180023
C 0.201764 -7.338548 -2.319023
C 0.874087 -6.271116 -3.045988
C 0.134561 -3.944593 -3.497975
C -0.142236 -5.322435 -3.499047
C -2.092785 -1.855469 -0.731436
C -0.882787 -2.998987 -3.043959
C -2.142516 -3.463937 -2.618834
C -2.762334 -2.881169 -1.438915
C -3.432695 -3.946996 -0.705819
C -3.432695 -3.946996 0.705819
C -1.444951 -5.807324 -3.050453
C -2.427860 -4.895773 -2.618744
C -3.230264 -5.194421 -1.436454
C -3.024781 -6.394859 -0.730312
C -2.009203 -7.340788 -1.179353
C -1.231213 -7.048729 -2.317774
C -3.024781 -6.394859 0.730312
C -2.009203 -7.340788 1.179353
C -1.380584 -7.924119 0.000000
C 0.003066 -8.200416 0.000000
C -1.231213 -7.048729 2.317774
C -3.230264 -5.194421 1.436454 \\

(C$_{60}$)$_2^+$, IV
E = -4568.590569
c 3.137179 -0.787591 3.034961
c 7.559720 -1.456992 1.810960
c 6.716640 -1.037962 2.857703
c 6.528382 0.389573 3.118593
c 7.193152 1.348133 2.326256
c 8.059454 0.914192 1.239869
c 8.239302 -0.465050 0.985252
c 5.442726 -1.707715 3.080356
c 4.466143 -0.696725 3.481477
c 5.141193 0.599250 3.504269
c 5.053022 -2.774825 2.245900
c 3.674024 -2.869170 1.780867
c 2.732179 -1.894050 2.168668
c 2.432718 0.415762 2.594483
c 1.602078 0.055049 1.456834
c 1.785380 -1.375405 1.193243
c 3.693161 -3.363844 0.406895
c 2.770714 -2.866260 -0.533229
c 1.800929 -1.852643 -0.131394
c 1.648265 -0.915707 -1.238526
c 1.468325 0.467615 -0.983602
c 1.444642 0.958040 0.384827
c 2.106317 2.258116 0.411299
c 2.917290 2.604114 1.507416
c 3.084478 1.666892 2.617110
c 4.461300 1.760416 3.081069
c 5.086136 -3.573553 0.023052
c 5.926741 -3.210455 1.160228
c 7.157783 -2.562354 0.946662
c 7.594464 -2.253110 -0.411918
c 6.783747 -2.601737 -1.507449
c 5.507330 -3.274883 -1.287069
c 3.208243 -2.556159 -1.892040
c 4.551135 -2.754921 -2.260030
c 2.507295 -1.350313 -2.327375
c 5.238579 -1.758293 -3.079754
c 4.558413 -0.597869 -3.505483
c 3.172669 -0.388822 -3.117634
c 2.139891 1.462593 -1.813062
c 2.983954 1.040404 -2.857406
c 2.540660 2.568517 -0.948662
c 4.257686 1.710076 -3.082413
c 4.647993 2.775612 -2.246376
c 3.774429 3.212459 -1.160844
c 4.194389 3.276831 1.287810
c 4.615715 3.573647 -0.022978
c 5.149354 2.755674 2.260778
c 6.491330 2.552306 1.890297
c 6.008173 3.362639 -0.407163
c 6.929027 2.861593 0.532254
c 7.901296 1.849446 0.131483
c 7.918523 1.373973 -1.193284
c 6.967137 1.892930 -2.166841
c 6.027428 2.869374 -1.781243
c 8.102058 -0.053966 -1.456561
c 7.265943 -0.414497 -2.591477
c 6.561532 0.787917 -3.032731
c 5.233482 0.698308 -3.482510
c 6.615607 -1.665531 -2.616934
c 8.264855 -0.956128 -0.385364
c -3.128976 0.732536 3.043997
c -7.193342 -1.377991 2.308769
c -6.523944 -0.434628 3.115193
c -6.707881 0.997536 2.877613
c -7.551512 1.436294 1.839362
c -8.235927 0.460179 0.999063
c -8.060118 -0.923608 1.231040
c -5.136683 -0.655049 3.494878
c -4.457400 0.638852 3.491549
c -5.431348 1.659330 3.108514
c -4.461427 -1.811382 3.051641
c -3.084979 -1.714962 2.586759
c -2.429188 -0.465805 2.583007
c -2.721940 1.851608 2.194942
c -1.778808 1.345681 1.209465
c -1.599396 -0.089470 1.449687
c -2.923038 -2.634512 1.461654
c -2.112925 -2.273450 0.369699
c -1.446895 -0.975406 0.363010
c -1.471493 -0.462776 -0.997423
c -1.647657 0.925108 -1.229617
c -1.795298 1.844352 -0.107325
c -2.762329 2.867639 -0.490893
c -3.681386 3.352903 0.458901
c -3.661334 2.835937 1.824649
c -5.039678 2.738605 2.290674
c -4.202837 -3.299240 1.233553
c -5.154303 -2.790870 2.216638
c -6.496365 -2.577278 1.852118
c -6.937456 -2.862953 0.489989
c -6.019958 -3.351669 -0.459127
c -4.627508 -3.573510 -0.081042
c -2.550936 -2.560247 -0.994275
c -3.787169 -3.196666 -1.214503
c -2.148075 -1.441826 -1.841482
c -4.661313 -2.739363 -2.291103
c -4.269030 -1.661644 -3.110532
c -2.992703 -0.999961 -2.877286
c -2.507092 1.380231 -2.309867
c -3.177131 0.433895 -3.114206
c -3.203212 2.581187 -1.853806
c -4.562951 0.653692 -3.496060
c -5.238489 1.809244 -3.050317
c -4.546211 2.790133 -2.215884
c -5.074352 3.573473 0.081160
c -5.498889 3.297317 -1.232804
c -5.914008 3.194742 1.213922
c -7.147534 2.554191 0.992253
c -6.778008 2.632142 -1.461659
c -7.587812 2.268455 -0.370322
c -8.262460 0.973457 -0.363461
c -8.104674 0.088294 -1.449362
c -7.269452 0.464436 -2.579940
c -6.615115 1.713572 -2.586525
c -7.925243 -1.344346 -1.209498
c -6.977388 -1.850576 -2.193070
c -6.569717 -0.732950 -3.041743
c -5.242235 -0.640428 -3.492539
c -6.040126 -2.836157 -1.825002
c -7.907054 -1.841215 0.107452 \\

C$_{60}^{2+}$
E = -2283.758106
6 -1.348501 -0.635421 -3.213872
6 -0.522061 -1.773184 -3.026821
6 0.669532 0.818522 -3.375552
6 -0.745921 0.676410 -3.390697
6 2.525487 1.470315 -2.076962
6 0.884641 2.781431 2.056320
6 2.059991 2.616768 1.285005
6 2.921909 1.444633 1.486173
6 2.574900 0.475273 2.443140
6 1.348504 0.635423 3.213869
6 0.522057 1.773176 3.026815
6 2.046521 2.928211 -0.127038
6 2.898135 1.952763 -0.814956
6 3.438468 1.037498 0.187844
6 0.854004 3.391904 -0.734395
6 0.467567 2.887348 -2.034419
6 1.283127 1.940306 -2.694957
6 2.671206 0.051284 -2.386984
6 1.519568 -0.348385 -3.184620
6 -0.992889 2.741746 -2.051447
6 -1.587523 1.653591 -2.726705
6 0.933601 -1.627280 -3.012239
6 1.469822 -2.546559 -2.031596
6 2.571302 -2.153728 -1.236150
6 3.184134 -0.831830 -1.420317
6 3.579631 -0.332269 -0.111788
6 -1.500887 3.156945 -0.760062
6 -0.361638 3.564289 0.062480
6 -0.346699 3.265988 1.430801
6 -1.469824 2.546565 2.031597
6 -2.571305 2.153732 1.236152
6 -2.589005 2.464596 -0.177616
6 -2.723154 0.943901 -2.134177
6 -3.211866 1.339522 -0.881279
6 -2.574904 -0.475274 -2.443143
6 -3.579627 0.332269 0.111789
6 -3.438463 -1.037496 -0.187841
6 -2.921903 -1.444628 -1.486171
6 -0.884641 -2.781432 -2.056320
6 -2.059992 -2.616770 -1.285006
6 0.346700 -3.265985 -1.430801
6 -2.046521 -2.928211 0.127037
6 -0.854006 -3.391907 0.734396
6 0.361637 -3.564287 -0.062480
6 2.589004 -2.464596 0.177617
6 1.500887 -3.156944 0.760061
6 3.211867 -1.339522 0.881278
6 2.723153 -0.943902 2.134176
6 0.992889 -2.741750 2.051448
6 1.587521 -1.653590 2.726709
6 0.745917 -0.676413 3.390693
6 -0.669528 -0.818522 3.375557
6 -1.283126 -1.940307 2.694959
6 -0.467567 -2.887350 2.034420
6 -1.519570 0.348385 3.184616
6 -2.671207 -0.051286 2.386985
6 -2.525485 -1.470316 2.076962
6 -2.898134 -1.952766 0.814957
6 -3.184132 0.831831 1.420318
6 -0.933605 1.627282 3.012239 \\

(C$_{60}$)$_2^{2+}$, I
E = -4568.299504
C 7.421032 1.182296 1.431856
C 6.998494 2.330401 0.729471
C 6.998494 2.330401 -0.729471
C 7.421032 1.182296 -1.431856
C 7.870846 -0.000000 -0.703651
C 7.870846 -0.000000 0.703651
C 5.831195 3.080826 -1.182305
C 5.108549 2.640098 -2.309201
C 5.533014 1.441012 -3.015832
C 6.674015 0.732920 -2.597620
C 6.674015 -0.732920 -2.597620
C 5.533014 -1.441012 -3.015832
C 4.329486 -0.704164 -3.380403
C 4.329486 0.704164 -3.380403
C 3.168055 1.420939 -2.903260
C 2.003208 0.744732 -2.421692
C 2.003208 -0.744732 -2.421692
C 3.168055 -1.420939 -2.903260
C 5.108549 -2.640098 -2.309201
C 3.649335 -2.646476 -2.280790
C 3.649335 2.646476 -2.280790
C 2.969673 3.164561 -1.168775
C 1.744359 2.532795 -0.740147
C 1.210279 1.420713 -1.426809
C 0.000000 0.772896 -0.810258
C 1.210279 -1.420713 -1.426809
C 1.744359 -2.532795 -0.740147
C 2.969673 -3.164561 -1.168775
C 7.421032 -1.182296 -1.431856
C 6.998494 -2.330401 -0.729471
C 5.831195 -3.080826 -1.182305
C 5.119409 -3.560416 0.000000
C 3.717705 -3.606487 0.000000
C -0.000000 0.772896 0.810258
C 1.210279 1.420713 1.426809
C 1.744359 2.532795 0.740147
C 2.969673 3.164561 1.168775
C 3.717705 3.606487 0.000000
C 5.119409 3.560416 0.000000
C 5.831195 3.080826 1.182305
C 5.108549 2.640098 2.309201
C 3.649335 2.646476 2.280790
C 2.003208 0.744732 2.421692
C 3.168055 1.420939 2.903260
C 6.674015 0.732920 2.597620
C 5.533014 1.441012 3.015832
C 4.329486 0.704164 3.380403
C 1.210279 -1.420713 1.426809
C 1.744359 -2.532795 0.740147
C 2.969673 -3.164561 1.168775
C 7.421032 -1.182296 1.431856
C 6.998494 -2.330401 0.729471
C 5.831195 -3.080826 1.182305
C 5.108549 -2.640098 2.309201
C 3.649335 -2.646476 2.280790
C 2.003208 -0.744732 2.421692
C 3.168055 -1.420939 2.903260
C 4.329486 -0.704164 3.380403
C 5.533014 -1.441012 3.015832
C 6.674015 -0.732920 2.597620
C -1.210279 -1.420713 -1.426809
C -1.744359 -2.532795 -0.740147
C -1.744359 -2.532795 0.740147
C -1.210279 -1.420713 1.426809
C -0.000000 -0.772896 0.810258
C -0.000000 -0.772896 -0.810258
C -2.003208 -0.744732 2.421692
C -3.168055 -1.420939 2.903260
C -3.649335 -2.646476 2.280790
C -2.969673 -3.164561 1.168775
C -4.329486 -0.704164 3.380403
C -4.329486 0.704164 3.380403
C -3.168055 1.420939 2.903260
C -2.003208 0.744732 2.421692
C -1.210279 1.420713 1.426809
C -1.744359 2.532795 0.740147
C -2.969673 3.164561 1.168775
C -3.649335 2.646476 2.280790
C -1.744359 2.532795 -0.740147
C -2.969673 3.164561 -1.168775
C -3.717705 3.606487 0.000000
C -5.119409 3.560416 0.000000
C -5.831195 3.080826 -1.182305
C -5.108549 2.640098 -2.309201
C -3.649335 2.646476 -2.280790
C -3.168055 1.420939 -2.903260
C -2.003208 0.744732 -2.421692
C -1.210279 1.420713 -1.426809
C -5.831195 3.080826 1.182305
C -5.108549 2.640098 2.309201
C -5.533014 1.441012 3.015832
C -5.108549 -2.640098 2.309201
C -5.533014 -1.441012 3.015832
C -6.674015 -0.732920 2.597620
C -6.674015 0.732920 2.597620
C -5.831195 -3.080826 1.182305
C -5.119409 -3.560416 -0.000000
C -3.717705 -3.606487 -0.000000
C -2.969673 -3.164561 -1.168775
C -6.998494 -2.330401 0.729471
C -7.421032 -1.182296 1.431856
C -6.998494 2.330401 0.729471
C -7.421032 1.182296 1.431856
C -7.870846 -0.000000 0.703651
C -7.870846 0.000000 -0.703651
C -7.421032 -1.182296 -1.431856
C -6.998494 -2.330401 -0.729471
C -5.831195 -3.080826 -1.182305
C -5.108549 -2.640098 -2.309201
C -3.649335 -2.646476 -2.280790
C -7.421032 1.182296 -1.431856
C -6.998494 2.330401 -0.729471
C -6.674015 0.732920 -2.597620
C -6.674015 -0.732920 -2.597620
C -5.533014 -1.441012 -3.015832
C -4.329486 -0.704164 -3.380403
C -3.168055 -1.420939 -2.903260
C -2.003208 -0.744732 -2.421692
C -4.329486 0.704164 -3.380403
C -5.533014 1.441012 -3.015832 \\

(C$_{60}$)$_2^{2+}$, II
E = -4568.266501
6 -0.794764 -0.000002 0.796958
6 -0.794765 -0.000002 -0.796963
6 0.794764 -0.000002 0.796958
6 0.794765 -0.000002 -0.796964
6 1.938721 2.300944 0.737385
6 1.496826 1.183233 1.454337
6 -1.938721 2.300943 0.737386
6 -1.496826 1.183232 1.454337
6 3.842427 3.491618 -0.000001
6 7.599754 1.182381 1.432344
6 7.160247 2.324246 0.728084
6 7.160248 2.324245 -0.728082
6 7.599756 1.182381 -1.432341
6 8.050996 0.000002 -0.702179
6 8.050995 0.000002 0.702183
6 5.982070 3.057414 1.181679
6 5.248988 3.504690 -0.000000
6 5.982071 3.057413 -1.181678
6 5.284351 2.628768 2.325017
6 3.824538 2.622478 2.324812
6 3.115723 3.039098 1.180789
6 3.115725 3.039097 -1.180793
6 1.938722 2.300943 -0.737390
6 3.381527 1.441521 3.046799
6 2.244989 0.732684 2.594908
6 1.496827 1.183232 -1.454342
6 2.244993 0.732683 -2.594911
6 3.381532 1.441520 -3.046801
6 3.824541 2.622476 -2.324814
6 5.284354 2.628766 -2.325017
6 4.558381 0.711003 3.483225
6 5.737817 1.444561 3.049724
6 6.870427 0.734488 2.605729
6 6.870428 -0.734485 2.605729
6 5.737819 -1.444559 3.049725
6 4.558381 -0.711001 3.483225
6 2.244990 -0.732686 2.594909
6 3.381529 -1.441522 3.046800
6 1.496827 -1.183235 1.454338
6 3.824540 -2.622477 2.324814
6 3.115726 -3.039099 1.180791
6 1.938723 -2.300946 0.737386
6 1.496828 -1.183236 -1.454342
6 1.938724 -2.300947 -0.737389
6 2.244993 -0.732687 -2.594911
6 3.115728 -3.039100 -1.180791
6 3.824543 -2.622478 -2.324813
6 3.381533 -1.441523 -3.046800
6 4.558385 0.711001 -3.483224
6 4.558386 -0.711003 -3.483224
6 5.737822 1.444560 -3.049723
6 6.870431 0.734487 -2.605726
6 5.737823 -1.444560 -3.049722
6 6.870432 -0.734486 -2.605726
6 7.599757 -1.182378 -1.432340
6 7.160250 -2.324243 -0.728080
6 5.982074 -3.057412 -1.181677
6 5.284357 -2.628766 -2.325016
6 7.160249 -2.324242 0.728085
6 5.982073 -3.057412 1.181680
6 5.248992 -3.504689 0.000001
6 3.842431 -3.491619 0.000000
6 5.284354 -2.628766 2.325018
6 7.599755 -1.182377 1.432345
6 -3.842427 3.491618 0.000000
6 -7.599753 1.182381 1.432345
6 -7.160246 2.324245 0.728085
6 -7.160247 2.324246 -0.728080
6 -7.599755 1.182381 -1.432340
6 -8.050996 0.000002 -0.702179
6 -8.050995 0.000002 0.702183
6 -5.982069 3.057414 1.181680
6 -5.248988 3.504690 0.000001
6 -5.982071 3.057414 -1.181677
6 -5.284351 2.628767 2.325018
6 -3.824537 2.622477 2.324814
6 -3.115723 3.039098 1.180791
6 -3.115724 3.039098 -1.180791
6 -1.938721 2.300944 -0.737389
6 -3.381527 1.441520 3.046800
6 -2.244989 0.732683 2.594908
6 -1.496827 1.183233 -1.454341
6 -2.244993 0.732684 -2.594911
6 -3.381531 1.441521 -3.046800
6 -3.824541 2.622477 -2.324813
6 -5.284354 2.628767 -2.325016
6 -4.558381 0.711001 3.483225
6 -5.737817 1.444560 3.049725
6 -6.870427 0.734488 2.605729
6 -6.870428 -0.734486 2.605729
6 -5.737819 -1.444560 3.049724
6 -4.558381 -0.711003 3.483225
6 -2.244990 -0.732687 2.594908
6 -3.381529 -1.441523 3.046799
6 -1.496827 -1.183236 1.454337
6 -3.824540 -2.622478 2.324812
6 -3.115726 -3.039100 1.180789
6 -1.938723 -2.300947 0.737385
6 -1.496828 -1.183235 -1.454342
6 -1.938724 -2.300946 -0.737390
6 -2.244994 -0.732686 -2.594911
6 -3.115728 -3.039099 -1.180793
6 -3.824544 -2.622477 -2.324814
6 -3.381533 -1.441521 -3.046801
6 -4.558385 0.711003 -3.483224
6 -4.558386 -0.711001 -3.483225
6 -5.737821 1.444561 -3.049722
6 -6.870431 0.734489 -2.605725
6 -5.737823 -1.444558 -3.049723
6 -6.870432 -0.734485 -2.605726
6 -7.599757 -1.182377 -1.432341
6 -7.160251 -2.324242 -0.728082
6 -5.982075 -3.057412 -1.181678
6 -5.284357 -2.628765 -2.325017
6 -7.160249 -2.324243 0.728084
6 -5.982073 -3.057413 1.181679
6 -5.248992 -3.504689 -0.000000
6 -3.842431 -3.491618 -0.000001
6 -5.284354 -2.628767 2.325017
6 -7.599755 -1.182378 1.432344 \\

(C$_{60}$)$_2^{2+}$, III
E = -4568.259702
6 -0.691230 -1.342981 1.100079
6 0.014160 -0.817638 -0.128504
6 0.691230 1.342981 1.100079
6 -0.014160 0.817638 -0.128504
6 -1.387318 2.160412 2.134107
6 -0.005119 1.857995 2.239460
6 -2.598959 -2.224401 -0.179922
6 -2.012803 -1.889716 1.068510
6 -2.962305 3.880735 1.850413
6 -0.606338 7.849193 1.395511
6 -1.896079 7.285487 1.425880
6 -2.613774 7.033207 0.176277
6 -2.021547 7.365395 -1.059366
6 -0.684646 7.943858 -1.087212
6 0.011714 8.182070 0.118941
6 -2.187795 6.188484 2.334066
6 -3.097385 5.263650 1.654575
6 -3.363301 5.792085 0.319770
6 -1.183770 5.697180 3.195553
6 -1.041156 4.264605 3.386435
6 -1.918014 3.368874 2.733232
6 -3.082215 2.969687 0.712156
6 -2.117307 1.893019 0.888130
6 0.388637 3.959646 3.496679
6 0.897556 2.773503 2.944844
6 -1.451086 1.356725 -0.216989
6 -1.690758 1.891626 -1.528531
6 -2.635842 2.927839 -1.718972
6 -3.345416 3.476014 -0.573949
6 -3.484764 4.916249 -0.775480
6 1.123066 5.210300 3.356067
6 0.152851 6.285953 3.170188
6 0.435708 7.338113 2.282377
6 1.699779 7.358494 1.548721
6 2.635842 6.325888 1.728934
6 2.342119 5.229888 2.648532
6 2.155265 2.793664 2.214662
6 2.857966 3.999314 2.063199
6 2.012803 1.889716 1.068510
6 3.463305 4.337105 0.771328
6 3.352444 3.453492 -0.326408
6 2.598959 2.224401 -0.179922
6 0.608683 1.390217 -1.412367
6 1.885389 1.957330 -1.435244
6 -0.424113 1.912292 -2.263371
6 2.184699 3.054265 -2.345272
6 1.186771 3.548959 -3.204765
6 -0.145951 2.968186 -3.163915
6 -2.334964 4.022172 -2.629488
6 -1.111815 4.041958 -3.339322
6 -2.864544 5.253342 -2.052780
6 -2.146820 6.455931 -2.192193
6 -0.378097 5.293366 -3.496238
6 -0.883667 6.476423 -2.925711
6 0.015987 7.398289 -2.242265
6 1.390409 7.097846 -2.148454
6 1.916233 5.877105 -2.745213
6 1.046472 4.989111 -3.406002
6 2.113509 7.350170 -0.901720
6 3.072430 6.273029 -0.719607
6 2.955703 5.360878 -1.859305
6 3.091647 3.978101 -1.663938
6 3.334373 5.769704 0.572849
6 1.434047 7.881948 0.211027
6 -3.091647 -3.978101 -1.663938
6 -1.434047 -7.881948 0.211027
6 -2.113509 -7.350170 -0.901720
6 -1.390409 -7.097846 -2.148454
6 -0.015987 -7.398289 -2.242265
6 0.684646 -7.943858 -1.087212
6 -0.011714 -8.182070 0.118941
6 -3.072430 -6.273029 -0.719607
6 -2.955703 -5.360878 -1.859305
6 -1.916233 -5.877105 -2.745213
6 -3.334373 -5.769704 0.572849
6 -3.463305 -4.337105 0.771328
6 -3.352444 -3.453492 -0.326408
6 -2.184699 -3.054265 -2.345272
6 -1.885389 -1.957330 -1.435244
6 -2.857966 -3.999314 2.063199
6 -2.155265 -2.793664 2.214662
6 -0.608683 -1.390217 -1.412367
6 0.424113 -1.912292 -2.263371
6 0.145951 -2.968186 -3.163915
6 -1.186771 -3.548959 -3.204765
6 -1.046472 -4.989111 -3.406002
6 -2.342119 -5.229888 2.648532
6 -2.635842 -6.325888 1.728934
6 -1.699779 -7.358494 1.548721
6 -0.435708 -7.338113 2.282377
6 -0.152851 -6.285953 3.170188
6 -1.123066 -5.210300 3.356067
6 -0.897556 -2.773503 2.944844
6 -0.388637 -3.959646 3.496679
6 0.005119 -1.857995 2.239460
6 1.041156 -4.264605 3.386435
6 1.918014 -3.368874 2.733232
6 1.387318 -2.160412 2.134107
6 1.451086 -1.356725 -0.216989
6 2.117307 -1.893019 0.888130
6 1.690758 -1.891626 -1.528531
6 3.082215 -2.969687 0.712156
6 3.345416 -3.476014 -0.573949
6 2.635842 -2.927839 -1.718972
6 1.111815 -4.041958 -3.339322
6 2.334964 -4.022172 -2.629488
6 0.378097 -5.293366 -3.496238
6 0.883667 -6.476423 -2.925711
6 2.864544 -5.253342 -2.052780
6 2.146820 -6.455931 -2.192193
6 2.021547 -7.365395 -1.059366
6 2.613774 -7.033207 0.176277
6 3.363301 -5.792085 0.319770
6 3.484764 -4.916249 -0.775480
6 1.896079 -7.285487 1.425880
6 2.187795 -6.188484 2.334066
6 3.097385 -5.263650 1.654575
6 2.962305 -3.880735 1.850413
6 1.183770 -5.697180 3.195553
6 0.606338 -7.849193 1.395511 \\

(C$_{60}$)$_2^{2+}$, IV
E = -4568.255085
6 -1.407598 -0.654134 -0.760444
6 -1.557639 0.684083 -1.231947
6 1.559687 -0.689083 -1.234399
6 1.406585 0.648150 -0.761170
6 1.772717 -1.526951 1.086753
6 1.733348 -1.792948 -0.293262
6 -1.626890 0.137835 1.575323
6 -1.439693 -0.927254 0.660199
6 3.198867 -1.241434 2.941070
6 7.557427 -1.750320 1.454474
6 6.759776 -1.490785 2.583231
6 6.598949 -0.114994 3.066060
6 7.247674 0.953048 2.405855
6 8.059580 0.685901 1.234911
6 8.211505 -0.644389 0.763224
6 5.489184 -2.177807 2.752679
6 4.542446 -1.231294 3.340925
6 5.236799 0.043070 3.532604
6 5.053861 -3.100772 1.780383
6 3.657154 -3.110058 1.361569
6 2.744261 -2.199338 1.932948
6 2.497619 0.024945 2.716215
6 1.626447 -0.146594 1.573615
6 3.617477 -3.386176 -0.071708
6 2.667173 -2.741604 -0.886263
6 1.437621 0.919335 0.659866
6 2.113039 2.198903 0.862227
6 2.970782 2.360970 1.963083
6 3.169697 1.255726 2.903507
6 4.557963 1.265669 3.321555
6 4.992025 -3.545620 -0.538422
6 5.880167 -3.371187 0.607012
6 7.110974 -2.707470 0.447834
6 7.497374 -2.192129 -0.862357
6 6.641393 -2.359016 -1.963993
6 5.366763 -3.049234 -1.801518
6 3.058576 -2.225046 -2.196269
6 4.381632 -2.374375 -2.641662
6 2.361996 -0.956446 -2.408590
6 5.052379 -1.263718 -3.319704
6 4.373097 -0.041591 -3.534505
6 3.013556 0.115163 -3.063453
6 2.052656 1.757880 -1.455787
6 2.852420 1.493134 -2.581688
6 2.497257 2.715865 -0.448909
6 4.122391 2.180714 -2.755190
6 4.558836 3.100904 -1.780715
6 3.732580 3.372513 -0.607126
6 4.246301 3.051114 1.802622
6 4.621771 3.544876 0.538517
6 5.229393 2.374980 2.642839
6 6.550925 2.220133 2.193132
6 5.995914 3.384680 0.071275
6 6.943482 2.735279 0.884225
6 7.881867 1.787816 0.293265
6 7.841023 1.523258 -1.086504
6 6.865244 2.196300 -1.930214
6 5.955944 3.110247 -1.361879
6 7.987855 0.147097 -1.572891
6 7.110517 -0.025058 -2.712421
6 6.409571 1.240388 -2.938200
6 5.067462 1.233101 -3.343133
6 6.441572 -1.255515 -2.904731
6 8.180209 -0.916324 -0.660824
6 -3.197694 1.234047 2.943531
6 -7.250760 -0.951349 2.403446
6 -6.600171 0.114422 3.065424
6 -6.757897 1.491203 2.584360
6 -7.554504 1.753960 1.455597
6 -8.210518 0.650348 0.762534
6 -8.061531 -0.680904 1.232453
6 -5.238549 -0.047082 3.532331
6 -4.541478 1.226117 3.342732
6 -5.485987 2.175378 2.755352
6 -4.562129 -1.270788 3.319944
6 -3.173651 -1.263113 2.902539
6 -2.498947 -0.033462 2.717266
6 -2.740647 2.192397 1.936934
6 -1.770089 1.519167 1.090281
6 -2.976553 -2.367474 1.960689
6 -2.117921 -2.205681 0.860473
6 -1.729632 1.786954 -0.289355
6 -2.661166 2.738364 -0.881489
6 -3.610498 3.383796 -0.066491
6 -3.651405 3.105774 1.366380
6 -5.048315 3.098777 1.784519
6 -4.253413 -3.054747 1.798681
6 -5.235517 -2.377776 2.639407
6 -6.556510 -2.219571 2.189283
6 -6.949505 -2.732079 0.879455
6 -6.002907 -3.382312 0.066059
6 -4.629313 -3.546003 0.533730
6 -2.502657 -2.719987 -0.451544
6 -3.739236 -3.373869 -0.611256
6 -2.055663 -1.761474 -1.456865
6 -4.564391 -3.098877 -1.784844
6 -4.125602 -2.178241 -2.757854
6 -2.854329 -1.493509 -2.582794
6 -2.358837 0.954756 -2.406233
6 -3.012314 -0.114560 -3.062828
6 -3.052964 2.224450 -2.192433
6 -4.371348 0.045633 -3.534262
6 -5.048216 1.268846 -3.318135
6 -4.375508 2.377156 -2.638259
6 -4.984497 3.546741 -0.533632
6 -5.359647 3.052840 -1.797584
6 -5.873529 3.372559 0.611150
6 -7.105623 2.711595 0.450486
6 -6.635623 2.365499 -1.961609
6 -7.492470 2.198860 -0.860589
6 -8.178016 0.924197 -0.661131
6 -7.987396 -0.138352 -1.574567
6 -7.109196 0.033580 -2.713480
6 -6.437644 1.262909 -2.903787
6 -7.843641 -1.515484 -1.090017
6 -6.868849 -2.189350 -1.934194
6 -6.410744 -1.232985 -2.940660
6 -5.068433 -1.227887 -3.344948
6 -5.961692 -3.105951 -1.366691
6 -7.885666 -1.781872 0.289374 \\

(C$_{60}$)$_2^{3+}$, I
E = -4567.880294
C 7.410591 1.180772 1.430597
C 6.989075 2.333295 0.727887
C 6.989075 2.333295 -0.727887
C 7.410591 1.180772 -1.430597
C 7.863732 -0.000000 -0.703333
C 7.863732 -0.000000 0.703333
C 5.824754 3.086429 -1.181612
C 5.101495 2.643056 -2.307080
C 5.526259 1.440239 -3.013913
C 6.668253 0.732815 -2.598915
C 6.668253 -0.732815 -2.598915
C 5.526259 -1.440239 -3.013913
C 4.324734 -0.704273 -3.381519
C 4.324734 0.704273 -3.381519
C 3.164466 1.420279 -2.900452
C 2.000293 0.744563 -2.421303
C 2.000293 -0.744563 -2.421303
C 3.164466 -1.420279 -2.900452
C 5.101495 -2.643056 -2.307080
C 3.645698 -2.649866 -2.278214
C 3.645698 2.649866 -2.278214
C 2.964674 3.172679 -1.168538
C 1.742246 2.537980 -0.738587
C 1.207569 1.421088 -1.425867
C 0.000000 0.772423 -0.808691
C 1.207569 -1.421088 -1.425867
C 1.742246 -2.537980 -0.738587
C 2.964674 -3.172679 -1.168538
C 7.410591 -1.180772 -1.430597
C 6.989075 -2.333295 -0.727887
C 5.824754 -3.086429 -1.181612
C 5.113690 -3.568733 0.000000
C 3.712298 -3.616470 0.000000
C -0.000000 0.772423 0.808691
C 1.207569 1.421088 1.425867
C 1.742246 2.537980 0.738587
C 2.964674 3.172679 1.168538
C 3.712298 3.616470 0.000000
C 5.113690 3.568733 0.000000
C 5.824754 3.086429 1.181612
C 5.101495 2.643056 2.307080
C 3.645698 2.649866 2.278214
C 2.000293 0.744563 2.421303
C 3.164466 1.420279 2.900452
C 6.668253 0.732815 2.598915
C 5.526259 1.440239 3.013913
C 4.324734 0.704273 3.381519
C 1.207569 -1.421088 1.425867
C 1.742246 -2.537980 0.738587
C 2.964674 -3.172679 1.168538
C 7.410591 -1.180772 1.430597
C 6.989075 -2.333295 0.727887
C 5.824754 -3.086429 1.181612
C 5.101495 -2.643056 2.307080
C 3.645698 -2.649866 2.278214
C 2.000293 -0.744563 2.421303
C 3.164466 -1.420279 2.900452
C 4.324734 -0.704273 3.381519
C 5.526259 -1.440239 3.013913
C 6.668253 -0.732815 2.598915
C -1.207569 -1.421088 -1.425867
C -1.742246 -2.537980 -0.738587
C -1.742246 -2.537980 0.738587
C -1.207569 -1.421088 1.425867
C -0.000000 -0.772423 0.808691
C -0.000000 -0.772423 -0.808691
C -2.000293 -0.744563 2.421303
C -3.164466 -1.420279 2.900452
C -3.645698 -2.649866 2.278214
C -2.964674 -3.172679 1.168538
C -4.324734 -0.704273 3.381519
C -4.324734 0.704273 3.381519
C -3.164466 1.420279 2.900452
C -2.000293 0.744563 2.421303
C -1.207569 1.421088 1.425867
C -1.742246 2.537980 0.738587
C -2.964674 3.172679 1.168538
C -3.645698 2.649866 2.278214
C -1.742246 2.537980 -0.738587
C -2.964674 3.172679 -1.168538
C -3.712298 3.616470 0.000000
C -5.113690 3.568733 0.000000
C -5.824754 3.086429 -1.181612
C -5.101495 2.643056 -2.307080
C -3.645698 2.649866 -2.278214
C -3.164466 1.420279 -2.900452
C -2.000293 0.744563 -2.421303
C -1.207569 1.421088 -1.425867
C -5.824754 3.086429 1.181612
C -5.101495 2.643056 2.307080
C -5.526259 1.440239 3.013913
C -5.101495 -2.643056 2.307080
C -5.526259 -1.440239 3.013913
C -6.668253 -0.732815 2.598915
C -6.668253 0.732815 2.598915
C -5.824754 -3.086429 1.181612
C -5.113690 -3.568733 -0.000000
C -3.712298 -3.616470 -0.000000
C -2.964674 -3.172679 -1.168538
C -6.989075 -2.333295 0.727887
C -7.410591 -1.180772 1.430597
C -6.989075 2.333295 0.727887
C -7.410591 1.180772 1.430597
C -7.863732 -0.000000 0.703333
C -7.863732 0.000000 -0.703333
C -7.410591 -1.180772 -1.430597
C -6.989075 -2.333295 -0.727887
C -5.824754 -3.086429 -1.181612
C -5.101495 -2.643056 -2.307080
C -3.645698 -2.649866 -2.278214
C -7.410591 1.180772 -1.430597
C -6.989075 2.333295 -0.727887
C -6.668253 0.732815 -2.598915
C -6.668253 -0.732815 -2.598915
C -5.526259 -1.440239 -3.013913
C -4.324734 -0.704273 -3.381519
C -3.164466 -1.420279 -2.900452
C -2.000293 -0.744563 -2.421303
C -4.324734 0.704273 -3.381519
C -5.526259 1.440239 -3.013913 \\

(C$_{60}$)$_2^{3+}$, II
E = -4567.845342
6 -0.797230 -0.000529 0.795844
6 -0.797230 -0.000529 -0.795845
6 0.797230 -0.000529 0.795844
6 0.797230 -0.000529 -0.795845
6 1.945753 2.303491 0.735007
6 1.500900 1.183069 1.453158
6 -1.945753 2.303491 0.735007
6 -1.500900 1.183069 1.453158
6 3.846730 3.501368 -0.000000
6 7.604531 1.184041 1.430708
6 7.164202 2.328003 0.726275
6 7.164202 2.328002 -0.726275
6 7.604530 1.184041 -1.430708
6 8.057605 0.000467 -0.700936
6 8.057605 0.000467 0.700936
6 5.985798 3.064202 1.181101
6 5.253141 3.513283 -0.000000
6 5.985798 3.064201 -1.181101
6 5.289603 2.636188 2.324058
6 3.829084 2.628645 2.322361
6 3.120866 3.046335 1.180055
6 3.120866 3.046335 -1.180055
6 1.945753 2.303491 -0.735008
6 3.386509 1.444627 3.041019
6 2.249752 0.733674 2.589265
6 1.500900 1.183069 -1.453159
6 2.249753 0.733674 -2.589265
6 3.386509 1.444626 -3.041018
6 3.829085 2.628645 -2.322361
6 5.289604 2.636187 -2.324058
6 4.563141 0.714591 3.468413
6 5.741947 1.449310 3.044548
6 6.873176 0.737100 2.598201
6 6.873430 -0.736510 2.598253
6 5.742440 -1.449011 3.044568
6 4.563344 -0.714571 3.468399
6 2.249987 -0.734268 2.589213
6 3.387009 -1.444897 3.040969
6 1.501230 -1.183881 1.453171
6 3.829929 -2.628841 2.322332
6 3.121859 -3.046690 1.180038
6 1.946504 -2.304141 0.734965
6 1.501230 -1.183881 -1.453172
6 1.946505 -2.304141 -0.734966
6 2.249988 -0.734268 -2.589213
6 3.121859 -3.046689 -1.180039
6 3.829929 -2.628840 -2.322331
6 3.387009 -1.444897 -3.040969
6 4.563142 0.714590 -3.468411
6 4.563344 -0.714571 -3.468398
6 5.741947 1.449310 -3.044547
6 6.873176 0.737100 -2.598201
6 5.742439 -1.449011 -3.044567
6 6.873429 -0.736509 -2.598252
6 7.604922 -1.183251 -1.430722
6 7.164963 -2.327340 -0.726320
6 5.986803 -3.063832 -1.181117
6 5.290446 -2.635988 -2.324072
6 7.164963 -2.327340 0.726320
6 5.986803 -3.063832 1.181117
6 5.254248 -3.513087 -0.000000
6 3.847837 -3.501551 -0.000000
6 5.290446 -2.635989 2.324072
6 7.604922 -1.183251 1.430722
6 -3.846730 3.501368 -0.000000
6 -7.604531 1.184041 1.430708
6 -7.164202 2.328002 0.726276
6 -7.164202 2.328003 -0.726275
6 -7.604531 1.184041 -1.430707
6 -8.057605 0.000467 -0.700936
6 -8.057605 0.000467 0.700937
6 -5.985797 3.064202 1.181101
6 -5.253141 3.513283 0.000000
6 -5.985798 3.064202 -1.181101
6 -5.289603 2.636187 2.324058
6 -3.829084 2.628645 2.322361
6 -3.120866 3.046335 1.180055
6 -3.120866 3.046335 -1.180055
6 -1.945753 2.303491 -0.735008
6 -3.386509 1.444627 3.041018
6 -2.249753 0.733674 2.589265
6 -1.500900 1.183069 -1.453159
6 -2.249753 0.733674 -2.589266
6 -3.386509 1.444627 -3.041018
6 -3.829085 2.628645 -2.322362
6 -5.289604 2.636188 -2.324058
6 -4.563141 0.714591 3.468412
6 -5.741947 1.449310 3.044548
6 -6.873176 0.737100 2.598201
6 -6.873429 -0.736509 2.598253
6 -5.742439 -1.449011 3.044568
6 -4.563344 -0.714571 3.468398
6 -2.249987 -0.734268 2.589213
6 -3.387009 -1.444897 3.040969
6 -1.501230 -1.183881 1.453171
6 -3.829929 -2.628841 2.322331
6 -3.121859 -3.046690 1.180038
6 -1.946505 -2.304141 0.734965
6 -1.501230 -1.183881 -1.453172
6 -1.946505 -2.304141 -0.734966
6 -2.249988 -0.734268 -2.589213
6 -3.121859 -3.046690 -1.180039
6 -3.829929 -2.628841 -2.322332
6 -3.387009 -1.444897 -3.040969
6 -4.563142 0.714591 -3.468412
6 -4.563344 -0.714571 -3.468399
6 -5.741948 1.449310 -3.044548
6 -6.873176 0.737100 -2.598201
6 -5.742440 -1.449011 -3.044568
6 -6.873430 -0.736510 -2.598252
6 -7.604922 -1.183251 -1.430722
6 -7.164963 -2.327340 -0.726319
6 -5.986803 -3.063832 -1.181117
6 -5.290447 -2.635988 -2.324072
6 -7.164963 -2.327340 0.726320
6 -5.986803 -3.063832 1.181117
6 -5.254248 -3.513087 0.000000
6 -3.847837 -3.501551 -0.000000
6 -5.290446 -2.635988 2.324072
6 -7.604922 -1.183251 1.430723 \\

(C$_{60}$)$_2^{3+}$, III
E = -4567.839188
6 -0.615171 -1.387177 1.071225
6 0.067815 -0.815370 -0.149860
6 0.615171 1.387177 1.071225
6 -0.067815 0.815370 -0.149860
6 -1.502799 2.063814 2.140372
6 -0.099301 1.855657 2.222446
6 -2.442965 -2.397326 -0.231186
6 -1.893108 -2.021946 1.024705
6 -3.193975 3.672529 1.889386
6 -1.113980 7.789905 1.424246
6 -2.363907 7.142773 1.469750
6 -3.077587 6.846766 0.228063
6 -2.522394 7.220444 -1.015160
6 -1.232012 7.885074 -1.057313
6 -0.537729 8.165606 0.140841
6 -2.569156 6.025462 2.375844
6 -3.423489 5.043949 1.702892
6 -3.738242 5.558751 0.374144
6 -1.521836 5.598684 3.221078
6 -1.280216 4.179679 3.398818
6 -2.105579 3.228389 2.753964
6 -3.265085 2.761392 0.746269
6 -2.226295 1.755939 0.902548
6 0.169857 3.969080 3.489231
6 0.749677 2.821563 2.924619
6 -1.537534 1.267324 -0.216962
6 -1.831030 1.786652 -1.521147
6 -2.845670 2.760539 -1.692889
6 -3.577263 3.255043 -0.534726
6 -3.813998 4.679859 -0.726778
6 0.816274 5.264657 3.346569
6 -0.227125 6.275540 3.180590
6 -0.027158 7.347005 2.296241
6 1.221810 7.451389 1.544131
6 2.226295 6.481912 1.704226
6 2.022744 5.368113 2.623362
6 1.991951 2.929240 2.177108
6 2.610575 4.177349 2.023616
6 1.893108 2.021946 1.024705
6 3.172400 4.560485 0.722425
6 3.107088 3.673782 -0.381155
6 2.442965 2.397326 -0.231186
6 0.498063 1.433490 -1.440370
6 1.733510 2.085730 -1.478367
6 -0.579519 1.890912 -2.275235
6 1.946461 3.203900 -2.387023
6 0.905605 3.634326 -3.229988
6 -0.384965 2.966346 -3.172550
6 -2.634862 3.872016 -2.603787
6 -1.423892 3.975568 -3.329346
6 -3.238057 5.062738 -2.013848
6 -2.603437 6.307902 -2.155308
6 -0.779035 5.271574 -3.489431
6 -1.355621 6.415462 -2.906351
6 -0.510257 7.385815 -2.224659
6 0.885977 7.180901 -2.155184
6 1.485612 6.004419 -2.768463
6 0.665939 5.062182 -3.419025
6 1.603431 7.471777 -0.918144
6 2.637019 6.467173 -0.755683
6 2.569156 5.555236 -1.900269
6 2.798864 4.183977 -1.713370
6 2.946561 5.976942 0.532700
6 0.899938 7.952081 0.212166
6 -2.798864 -4.183977 -1.713370
6 -0.899938 -7.952081 0.212166
6 -1.603431 -7.471777 -0.918144
6 -0.885977 -7.180901 -2.155184
6 0.510257 -7.385815 -2.224659
6 1.232012 -7.885074 -1.057313
6 0.537729 -8.165606 0.140841
6 -2.637019 -6.467173 -0.755683
6 -2.569156 -5.555236 -1.900269
6 -1.485612 -6.004419 -2.768463
6 -2.946561 -5.976942 0.532700
6 -3.172400 -4.560485 0.722425
6 -3.107088 -3.673782 -0.381155
6 -1.946461 -3.203900 -2.387023
6 -1.733510 -2.085730 -1.478367
6 -2.610575 -4.177349 2.023616
6 -1.991951 -2.929240 2.177108
6 -0.498063 -1.433490 -1.440370
6 0.579519 -1.890912 -2.275235
6 0.384965 -2.966346 -3.172550
6 -0.905605 -3.634326 -3.229988
6 -0.665939 -5.062182 -3.419025
6 -2.022744 -5.368113 2.623362
6 -2.226295 -6.481912 1.704226
6 -1.221810 -7.451389 1.544131
6 0.027158 -7.347005 2.296241
6 0.227125 -6.275540 3.180590
6 -0.816274 -5.264657 3.346569
6 -0.749677 -2.821563 2.924619
6 -0.169857 -3.969080 3.489231
6 0.099301 -1.855657 2.222446
6 1.280216 -4.179679 3.398818
6 2.105579 -3.228389 2.753964
6 1.502799 -2.063814 2.140372
6 1.537534 -1.267324 -0.216962
6 2.226295 -1.755939 0.902548
6 1.831030 -1.786652 -1.521147
6 3.265085 -2.761392 0.746269
6 3.577263 -3.255043 -0.534726
6 2.845670 -2.760539 -1.692889
6 1.423892 -3.975568 -3.329346
6 2.634862 -3.872016 -2.603787
6 0.779035 -5.271574 -3.489431
6 1.355621 -6.415462 -2.906351
6 3.238057 -5.062738 -2.013848
6 2.603437 -6.307902 -2.155308
6 2.522394 -7.220444 -1.015160
6 3.077587 -6.846766 0.228063
6 3.738242 -5.558751 0.374144
6 3.813998 -4.679859 -0.726778
6 2.363907 -7.142773 1.469750
6 2.569156 -6.025462 2.375844
6 3.423489 -5.043949 1.702892
6 3.193975 -3.672529 1.889386
6 1.521836 -5.598684 3.221078
6 1.113980 -7.789905 1.424246 \\

(C$_{60}$)$_2^{3+}$, IV
E = -4567.836486
6 -1.492075 -0.673220 -0.982498
6 -1.579131 0.733391 -1.201795
6 1.578640 -0.731734 -1.201506
6 1.492434 0.674969 -0.982422
6 1.702465 -1.130520 1.249445
6 1.679963 -1.645530 -0.068335
6 -1.615323 -0.315094 1.474552
6 -1.507813 -1.201746 0.373714
6 3.089180 -0.586973 3.071137
6 7.439580 -1.591784 1.839235
6 6.634722 -1.086115 2.878556
6 6.545924 0.361943 3.102163
6 7.268472 1.252877 2.280202
6 8.082969 0.729559 1.200127
6 8.167006 -0.671603 0.982051
6 5.324628 -1.659606 3.125711
6 4.421260 -0.574192 3.506211
6 5.184053 0.674959 3.488043
6 4.858181 -2.724372 2.321343
6 3.478888 -2.735213 1.866960
6 2.605862 -1.687856 2.235608
6 2.466210 0.650461 2.597932
6 1.615584 0.317197 1.474667
6 3.455254 -3.269872 0.503726
6 2.561090 -2.738143 -0.446879
6 1.508498 1.203726 0.373686
6 2.249000 2.458509 0.361794
6 3.091548 2.775240 1.443318
6 3.204577 1.855884 2.579075
6 4.581796 1.871052 3.035429
6 4.823590 -3.588068 0.121096
6 5.694582 -3.253440 1.244998
6 6.960616 -2.695826 1.007510
6 7.407452 -2.453429 -0.362627
6 6.568751 -2.775012 -1.444732
6 5.254528 -3.351709 -1.200702
6 3.007219 -2.497097 -1.818370
6 4.328314 -2.795073 -2.184996
6 2.387961 -1.255564 -2.283183
6 5.077397 -1.868224 -3.034332
6 4.474893 -0.673260 -3.490258
6 3.114827 -0.361023 -3.100037
6 2.216570 1.597348 -1.840733
6 3.025770 1.087932 -2.876195
6 2.694876 2.700639 -1.008593
6 4.334657 1.662079 -3.127688
6 4.801657 2.723851 -2.320444
6 3.965387 3.252158 -1.243886
6 4.405863 3.354240 1.201418
6 4.836959 3.586821 -0.120471
6 5.330581 2.797770 2.186254
6 6.649252 2.493778 1.815593
6 6.206353 3.270197 -0.504267
6 7.096414 2.733993 0.445190
6 7.982372 1.642401 0.067994
6 7.959048 1.129002 -1.248074
6 7.051071 1.687183 -2.232217
6 6.182420 2.736736 -1.867059
6 8.043287 -0.315616 -1.472850
6 7.189548 -0.649246 -2.594316
6 6.566700 0.587742 -3.068310
6 5.237102 0.576328 -3.509248
6 6.455381 -1.855772 -2.580326
6 8.150360 -1.200735 -0.373429
6 -3.089351 0.588498 3.070945
6 -7.267638 -1.253881 2.280544
6 -6.545552 -0.362393 3.102310
6 -6.635196 1.085572 2.878469
6 -7.440394 1.590597 1.839101
6 -8.167348 0.669856 0.982118
6 -8.082495 -0.731215 1.200424
6 -5.183481 -0.674564 3.488179
6 -4.421404 0.575024 3.506089
6 -5.325415 1.659852 3.125449
6 -4.580570 -1.870389 3.035740
6 -3.203382 -1.854507 2.579308
6 -2.465702 -0.648658 2.597919
6 -2.606704 1.689515 2.235204
6 -1.703034 1.132532 1.249092
6 -3.089891 -2.773991 1.443701
6 -2.247585 -2.456955 0.362077
6 -1.680891 1.647328 -0.068781
6 -2.562678 2.739366 -0.447464
6 -3.457099 3.270745 0.503096
6 -3.480355 2.736307 1.866423
6 -4.859616 2.724749 2.320874
6 -4.403885 -3.353789 1.201968
6 -5.328871 -2.797684 2.186761
6 -6.647737 -2.494514 1.816119
6 -7.094827 -2.735213 0.445778
6 -6.204507 -3.271068 -0.503636
6 -4.834915 -3.586847 -0.119859
6 -2.693390 -2.699578 -1.008243
6 -3.963593 -3.251876 -1.243377
6 -2.215743 -1.596166 -1.840607
6 -4.800223 -2.724233 -2.319976
6 -4.333869 -1.662332 -3.127423
6 -3.025291 -1.087397 -2.876111
6 -2.388808 1.256563 -2.283513
6 -3.115203 0.361465 -3.100179
6 -3.008740 2.497834 -1.818891
6 -4.475465 0.672858 -3.490390
6 -5.078622 1.867557 -3.034638
6 -4.330024 2.794986 -2.185497
6 -4.825634 3.588096 0.120485
6 -5.256508 3.351262 -1.201251
6 -5.696376 3.253157 1.244486
6 -6.962097 2.694767 1.007157
6 -6.570411 2.773768 -1.445113
6 -7.408872 2.451884 -0.362912
6 -8.151065 1.198765 -0.373460
6 -8.043549 0.313519 -1.472737
6 -7.190056 0.647446 -2.594304
6 -6.456575 1.854393 -2.580556
6 -7.958473 -1.131009 -1.247721
6 -7.050228 -1.688841 -2.231813
6 -6.566529 -0.589267 -3.068119
6 -5.236957 -0.577166 -3.509125
6 -6.180954 -2.737833 -1.866520
6 -7.981427 -1.644192 0.068443 \\